# Linear instability of the lid-driven flow in a cubic cavity

Alexander Yu. Gelfgat

*School of Mechanical Engineering, Faculty of Engineering, Tel-Aviv University, Ramat Aviv, Tel-Aviv 69978, Israel*

**Abstract**

Primary instability of the lid-driven flow in a cube is studied by a linear stability approach. Two cases, in which the lid moves parallel to the cube sidewall or parallel to the diagonal plane, are considered. It is shown that Krylov vectors required for application of the Newton and Arnoldi iteration methods can be evaluated by the SIMPLE procedure. The finite volume grid is gradually refined from $100^3$ to $256^3$ nodes. The computations result in grid converging values of the critical Reynolds number and oscillation frequency that allow for Richardson extrapolation to the zero grid size. Three-dimensional flow and most unstable perturbations are visualized by a recently proposed approach that allows for a better insight in the flow patterns and appearance of the instability. New arguments regarding the assumption that the centrifugal mechanism triggers the instability are given for both cases.

## 1. Introduction

For several decades the lid-driven flow in a square cavity was, and continues to be, one of the most popular CFD benchmarks. The two-dimensional lid-driven flow is very well studied, and main results are summarized in the review papers [1,2]. Nevertheless, this problem continues to attract attention, as can be seen from several recent publications [3-5]. The three-dimensional lid-driven flow is substantially more computationally challenging. First 3D steady state flows at low and moderate Reynolds numbers were calculated in [6]. Much later, accurate benchmark-quality results for the 3D steady state at *Re*=1000 were reported in [7] and were validated experimentally in [8]. Comparison of the 3D steady states calculated by the finite volume method applied here with the pseudospectral calculations of [7] are reported in [9,10]. The two independent results coincided to within the third decimal place. The most recent results on the three-dimensional lid-driven cavity flow were recently reviewed in [2]. In the present study, basing on our results [9,10], we assume that steady state flows can be computed accurately for the Reynolds numbers below 3000, and focus on examination of their linear stability and finding the instability limit.

It is well known that computation of a primary steady–oscillatory transition is a more challenging problem, compared to the calculation of a steady flow only, since it requires also accurate computation of the most unstable perturbation mode represented by the leading eigenvector of the momentum and continuity equations linearized around the steady state, so that application of the eigenvalue analysis would be the most natural choice for such kind of problems. However, the size of the algebraic eigenvalue problem becomes so large that the task seems to be unaffordable. Thus, the instability studies for the lid-driven cavity in a cube [10-16] are carried out using the straightforward time integration, rather than the eigenvalue analysis. In several other studies [16,17], the leading eigenvalues were computed to determine whether the flow is stable or unstable, however no attempt to arrive at an accurate value of the critical Reynolds number was made. The first thorough linear stability analysis was performed only recently in [18].

In the present study we consider two different problems of the lid-driven flow in a cube. In the first, the classical one, the lid moves parallel to one of the side walls, so that it becomes a straightforward 3D extension of the famous benchmark of lid-driven flow in a square cavity. In the second problem, proposed in [19,20], we consider the lid moving parallel to the diagonal plane of the cube. The benchmark results for steady states of this flow can be found in [21]. Its steady-oscillatory transition was studied recently in [22], again by the straightforward integration in time.

Our primary goal is to apply the eigenvalue analysis for calculation of the critical Reynolds numbers, at which the steady flow bifurcates to an oscillatory state. The computations are carried out using gradually refined stretched grids, containing $100^3$, $150^3$, $200^3$, and $256^3$ finite volumes. In spite that pseudo-spectral and collocation methods may yield better accuracy for model problems in simple geometries, as those considered here, they are not applicable to most of the applied problems that involve more complicated flow regions and liquid-liquid interfaces. Besides a wide set of curved-boundary-fitted numerical methods, such problems can be treated by the fixed grid finite volume approach used below with the immersed boundary technique, as it was done recently in [23].

The critical Reynolds numbers are calculated together with the critical oscillation frequencies. These are defined by the imaginary part of the leading eigenvalue, whose real part crosses the imaginary axis at the stability limit. The most unstable infinitesimally small perturbation is defined by the eigenvector corresponding to the leading eigenvalue. The eigenvector spatial pattern allows us to compare with the results of previous studies that applied the straightforward integration in time, to visualize spatio-temporal behavior of the most unstable disturbance, and basing on this to speculate further about the reasons that cause the instability onset.

In the following we formulate the problem and briefly describe the numerical method, as well as our method of visualization of three-dimensional divergence-free velocity field [24,25]. The present numerical approach includes Newton iteration for calculation of the steady states and Arnoldi iteration for computation of leading eigenvalues. The Newton corrections are computed by the Krylov-subspace-iteration techniques BiCGstab(2) or GMRES. The SIMPLE procedure [26] is reformulated for evaluation of the Krylov needed for BiCGstab(2), GMRES and Arnoldi methods [27]. To the best of the author's knowledge the SIMPLE procedure is applied for this purpose for the first time. Then, in the "Results and discussion" section we revisit the stability problem of the 2D lid-driven flow in square cavity, and compare several previously published results, which exhibit a noticeable scatter, with the present one. We show that grid converging result can be obtained using the finite volume discretization applied in this study. We show also that the Richardson extrapolation to zero grid size improves the result in site of the corner discontinuities of the problem. Basing on the conclusions made for the 2D problem, we report the convergence of our three-dimensional stability results that are compared to those obtained using exactly the same spatial discretization, but by means of the straightforward integration in time [10,22]. This is followed by comparison with all the other previously published results.

To visualize the three-dimensional flows we apply the novel method of two-dimensional divergence-free projections described in [24,25]. Along with a better insight in the 3D flow structure, this technique helps us to define location of the center of the main circulation, which in its turn allows for a physically consistent evaluation of the Rayleigh instability criterion. The Bayly criterion [28], yielding a sufficient condition for instability of inviscid flow with closed streamlines, is also calculated using approach of [29]. This yields an additional argument in favor of centrifugal instability mechanism that was assumed to drive the instability in [22,30,44]. Note that the Rayleigh or Bayly criteria were not calculated in the above studies, however, since they hold only for inviscid flows, they provide only a heuristic indication of possibility of the centrifugal instability mechanism. Furthermore, using the same visualization technique, we arrived here at quite a new way of visualization of the onset of centrifugal instability. We expect that the centrifugal instability sets in as vortical motion in the planes orthogonal to a local "rotational" plane, as it happens, e.g., in the Taylor-Couette flow. Calculation of the divergence free velocity projections on these planes allows us to examine whether such vortical motion is observed. Its existence confirms the centrifugal instability mechanism, while absence of such motion shows that the instability sets in a different way. Thus, we confirm the centrifugal mechanism for the lid driven parallel to the wall, and deny it for the diagonally driven lid case.

## 2. Formulation of the problem

Flow in a cubic cavity $0 \leq x, y, z \leq L$, with the side of length $L$ is considered. The boundary, at $z = L$, moves with a constant velocity $U$ in an arbitrary direction inside the $(x, y, z = L)$ plane, while all other boundaries are stationary. The no-slip boundary conditions are applied on all the boundaries. The flow is described by the continuity and momentum equations

$$\nabla \cdot \boldsymbol{u} = 0, \qquad (1)$$

$$\frac{\partial \boldsymbol{u}}{\partial t} + (\boldsymbol{u} \cdot \nabla)\boldsymbol{u} = -\nabla p + \frac{1}{Re}\Delta \boldsymbol{u}, \qquad (2)$$

where dimensionless variables are velocity $\boldsymbol{u} = (u, v, w)$, pressure $p$, and time $t$. The equations are rendered dimensionless using the scales $L, U, L/U$ and $\rho U^2$ for length, velocity, time and pressure, respectively. Here $\rho$ is the fluid density. The Reynolds number is defined by $Re =$ $UL/\nu$, where $\nu$ is the kinematic viscosity. Assuming that the angle between the lid velocity and the $x$ direction is $\alpha$, the dimensionless boundary conditions read

$$u = v = w = 0 \quad \text{at} \ x = 0,1; y = 0,1; \text{ and } z = 0; \quad (3)$$

$$w = 0, \quad u = cos(\alpha), \ v = sin(\alpha) \ \text{at} \ z = 1. \quad (4)$$

In the following we call the borders $z = 0,1$ horizontal, lower and upper, respectively, the borders $x = 0,1$ vertical, and the borders $y = 0,1$ spanwise.

## 3. Numerical method

The governing equations were discretized by the finite volume method, using the same schemes and the same staggered grid stretching as in [9,10,24,31]. Details on the finite volume discretization can be found in [31]. For the initially uniformly distributed $N_x \times N_y \times N_z$ grid points, $x_i = ih_x, y_j = jh_y, z_k = kh_z$, $\{h_x, h_y, h_z\} = 1/\{N_x, N_y, N_z\}$, $\{i,j,k\} \in [0, \{N_x, N_y, N_z\}]$, $\{x, y, z\} \in [0,1]$, the stretching function is defined by

$$\{x_i, y_j, z_k\} \to \{x_i, y_j, z_k\} - 0.0975 sin[2\pi\{x_i, y_j, z_k\}], \quad (5)$$

so that the grid nodes are distributed identically in each spatial direction. Apparently, the stretched grid points are distributed symmetrically with respect to $\{x_i, y_j, z_k\} = 0.5$. Thus, the distance between the corner and the closest grid point, which is the smallest grid step, is equal to $h_x - 0.0975 sin(2\pi h_x) \approx 0.387 h_x = 0.387/N_x$. The largest grid step corresponds to the distance between the centerline $x = 0.5$ and the neighbor point, for which a similar calculation yields $\approx 1.387/N_x$. Thus, the refinement ratio, defined as a ratio of finer and coarser grid steps, for the smallest and the largest grid steps is equal to the ratio of the grid point numbers, as is defined in [32].

The steady state flows were calculated by the Newton method and the leading eigenvalue and eigenvector required for the linear stability analysis were computed by the Arnoldi method using either the ARPACK package [33] or the EB13 module of the HSL library [34]. The Arnoldi method itself is a version of the Krylov subspace iteration, while corrections of the Newton method were calculated using either restarted GMRES or BiCGstab(2), both basing on the Krylov subspaces [27]. Therefore, to make the whole computational process robust, one needs an efficient calculation of the Krylov basis vectors. The latter is discussed below.

The classical textbook definition of Krylov basis vectors uses an arbitrary vector $x$ and a non-singular matrix $A$, so that Krylov basis vectors are defined as $x, Ax, A^2 x, ..., A^n x, ...$ [27]. However, for linearized and discretized incompressible Navier-Stokes equations, evaluation of a direct matrix-vector product is not sufficient. The equation (2) linearized in the neighborhood of a steady state flow denoted by $\boldsymbol{U}$ and $P$, is

$$\frac{\partial \boldsymbol{u}}{\partial t} + (\boldsymbol{U} \cdot \nabla)\boldsymbol{u} + (\boldsymbol{u} \cdot \nabla)\boldsymbol{U} = -\nabla p + \frac{1}{Re}\Delta \boldsymbol{u} \qquad (6)$$

After the linearization, the continuity equation (1) and the boundary conditions (3) remain the same, while the boundary conditions (4) become homogeneous. Altogether, the linearized momentum equation, the continuity equation, and the (linearized, if required) homogeneous boundary conditions form the Jacobian operator $\mathfrak{J}$ that acts simultaneously on the velocity and pressure fields. Then the Newton method aimed to computation of a steady state at a certain Reynolds number can be described as follows:

1. Choose an initial guess $(\boldsymbol{U}, P)$;
2. Substitute $(\boldsymbol{U}, P)$ into Eqs (1)-(4) and compute the residual vector $\boldsymbol{F}$; If $\|\boldsymbol{F}\| < \varepsilon$ exit.
3. Solve $\mathfrak{J}\begin{bmatrix}\boldsymbol{u}\\p\end{bmatrix} = \boldsymbol{F}$
4. Make a new guess $\boldsymbol{U} \to \boldsymbol{U} - \boldsymbol{u}$, $P \to P - p$ and go to the step 2.

The eigenvalue problem associated with the linear stability analysis reads

$$\mathfrak{J}\begin{bmatrix}\boldsymbol{u}\\p\end{bmatrix} = \lambda \begin{bmatrix}\boldsymbol{u}\\p\end{bmatrix}. \qquad (7)$$

Note that the Jacobian matrix $\mathfrak{J}$ is the same for both the Newton iteration and the stability problem. The steady flow $(\boldsymbol{U}, P)$ is unstable if there exists at least one eigenvalue with a positive real part. In the following we call the eigenvalue with the maximal real part $\Lambda$, as well as the associated eigenvector, "leading". The instability sets in at the critical value of Reynolds number $Re = Re_{cr}$, at which $\Lambda$ crosses the imaginary axis, so that its real part turns from negative to positive. In a slightly supercritical regime the most unstable disturbance oscillates with the critical circular frequency $\omega_{cr} \approx Im(\Lambda)$ and the exponentially growing amplitude defined by the leading eigenvector.

Note that both the Newton method and the eigenvalue problem (7) treat the same Jacobian operator $\mathfrak{J}$. Assuming that the equations (1) – (4) and (6) are discretized by some numerical approach, the Jacobian operator reduces to a Jacobian matrix, which defines either the linear algebraic equation system of the Step 3 of the Newton method, or the eigenvalue problem (6). As mentioned, these two problems are treated here by the Krylov-subspace iteration methods. Namely, the linear algebraic equations system is solved by either BiCGstab(2) or GMRES(n), and the eigenvalue problem is treated by the Arnoldi iteration [27,33,34].

Assume that the equations (1) – (4) are discretized by some numerical approach, and $v^{(n)}, p^{(n)}$ form a current Krylov vector. The next vector, $v^{(n+1)}, p^{(n+1)}$, must satisfy simultaneously

$$v^{(n+1)} = -\nabla p^{(n+1)} + \frac{1}{Re}\Delta v^{(n)} - (U \cdot \nabla)v^{(n)} - (v^{(n)} \cdot \nabla)U, \qquad (8)$$

$$\nabla \cdot v^{(n+1)} = 0; \ +all\ the\ homogeneous\ boundary\ conditions \ . \qquad (9)$$

If the expressions (8) and (9) are satisfied, the Krylov vectors belong to a subspace of divergence-free vectors satisfying all the homogeneous boundary conditions of the linearized problem. Contrarily, if, say, any of the constraints (9) is not satisfied, the Krylov iterations will seek the solution in a noticeably "wider" space, which in most cases leads to the loss of convergence. The most common way to overcome this difficulty is application of the Stokes preconditioning, that can be computationally realized via carrying out the time steps of either full or linearized problems [35-37]. It was argued in [38] that the Stokes operator can be replaced by a more general one. This approach is effective for 2D problems, as well as for the stability problems with a periodic third direction, so that the base flow is two-dimensional. Applied to a fully three-dimensional problem, like one considered here, this approach exhibits a slowed down convergence, especially when the grids are refined.

It was noted in [35] and later in [38] that a correct result $v^{(n+1)}, p^{(n+1)}$ of the problem (7) and (8) can be interpreted as a projection of a vector

$$w^{(n+1)} = \frac{1}{Re}\Delta v^{(n)} - (U \cdot \nabla)v^{(n)} - (v^{(n)} \cdot \nabla)U \qquad (10)$$

onto the above subspace of divergent free vectors satisfying all the linear and homogeneous boundary conditions. In [38] we proposed to approximate this projection using divergent-free basis functions built as linear superpositions of the Chebyshev polynomials. The approximate projection can be used as initial guess for a more CPU time consuming numerical process. Here we apply a different and more effective approach, which recalls the idea of SIMPLE [26] iteration, and propose to compute the correct projection via the following algorithm.

Start with $v^{(n+1)} = w^{(n+1)}$ from Eq. (8), and $p^{(n+1)} = 0$

Repeat until $\|\varphi\| < \varepsilon$

1. Solve $\Delta\varphi = \nabla \cdot v^{(n+1)}; \ \left[\frac{\partial\varphi}{\partial n}\right]_\Gamma = 0$.
2. Correct $v^{(n+1)}: v^{(n+1)} \to v^{(n+1)} - \nabla\varphi, p^{(n+1)} = p^{(n+1)} + \varphi$
3. If boundary conditions for $v^{(n+1)}$ are not satisfied, introduce them by changing the boundary values of $v^{(n+1)}$ and go to stage 1.

Steps 1 and 2 of this algorithm are the Chorin projection [39] that yields divergence-free velocity field. At the boundaries, owing to the Newman boundary conditions for $\varphi$, this projection keeps the normal velocity component unchanged. However, it can alter the tangent component. Note also, that after calculation of $\boldsymbol{w}^{(n+1)}$ by Eq. (10), no boundary conditions are generally kept. At the Step 3, the boundary conditions are restored by alteration of $\boldsymbol{v}^{(n+1)}$ boundary values. Obviously, this alters the divergence in the nodes adjacent to the boundaries, so that steps 1 and 2 should be repeated. It is easy to see that if the iterations converge, then the resulting fields $\boldsymbol{v}^{(n+1)}$ and $p^{(n+1)}$ are those required by (8) and (9). Note that the above iterative procedure makes no assumptions regarding problem or numerical discretization scheme, except the assumption of incompressibity.

The above algorithm was applied in the following computations for generation of the Krylov basis for GMRES, BiCGstab(2) and Arnoldi methods. The GMRES method was restarted after each 100 Krylov vectors computed. In the cases when GMRES saturated, its last approximation was supplied to the BiCGstab(2) as initial guess, which finally yielded a converged solution.

A special attention was paid to an accurate solution of the Poisson equation at the Step 1, for which an analytic TPT (Tensor Product with Thomas algorithm) method [40] was applied. This method combines the Thomas algorithm in one spatial dimension and the eigenvalue decomposition of the second derivative operators in the two other directions, as proposed in [41]. Note, that since the method is analytic, the solution $\varphi$ is obtained to within computer accuracy. Thus, in the computations below, after the correction of Step 2 was completed, the maximal absolute values of the discretized divergence of $\boldsymbol{v}^{(n+1)}$ remained below $10^{-8}$. When calculating on the finest grid, the TPT algorithm was replaced by the TPF (Tensor Product Factorization), which does not suffer from loss of accuracy of the Thomas algorithm when Neumann boundary value problem is considered [40]. With the TPF algorithm the maximal absolute value of divergence remains below $10^{-12}$. As shown in [40] both methods consume the order of $N^2$ ($N$ is the number of unknowns) operations, while TPT method is faster.

Both TPF and TPT methods were parallelized using either OpenMP, MPI, or hybrid MPI/OpenMP tools. Parallelization by the OpenMP is most effective for shared memory architectures and strongly depends on effective usage of CPU cache memory. Parallelization using the MPI directives suffers mainly from the necessity of matrix row/column swaps that cause too much communication. Most of the calculations reported below were performed of HP Xeon E5-2630 workstation with 24 CPU threads. Computational experiments on different platforms showed that using 32 CPUs we can speed up the code 6 – 7 times. With adding more CPUs the computational time quickly saturates. One characteristic run of the TPT algorithm consumed about ≈0.12 sec per CPU thread for the $100^3$ grid and ≈1.5 sec per CPU thread for the $256^3$ grid. Further optimization of the codes is clearly needed, but is beyond the scope of present study.

In most of the previous studies (e.g., [35-38,42]) the Arnoldi iteration was applied in the shift-and-invert mode, so that the eigenvalue having maximal absolute value was closest to the shift. If the leading eigenvalue, i.e., the eigenvalue with the largest real part, is well estimated, the appropriate shift can be easily applied. However, if there is no estimation of the leading eigenvalue, several calculations with different shifts must be performed in order to find the leading eigenvalue. With the method proposed above to calculate the Krylov vectors, the Arnoldi method can be run in a mode allowing for calculation of the leading eigenvalue in a single run (see [33,34] for the details). This is an obvious advantage, which allowed us to complete the computations in all the cases reported below. It should be noticed, however, that in spite of calculating of the Krylov vectors using the above algorithm is fast, the convergence of the Arnoldi iteration in this mode can be very slow.

For all grids considered the converged steady state is obtained after 6 – 8 Newton iterations. At the same time computation of the Newton correction at each iteration becomes noticeably more difficult with the grid refinement. Thus, for $100^3$ grid, the BiCGstab(2) process converges within 1000 iterations. Starting from $150^3$ grid, the BiCGstab(2) iterations do not converge. To reach the convergence, we apply the GMRES with 100 Krylov vector restarting until it saturates. Then we use the saturated results as an initial guess for BiCGstab(2) and iterate until convergence. The GMRES method saturates after 2 – 7 restarts for $150^3$, 5 – 20 restarts for $200^3$, and 10 – 40 restarts for $256^3$ grid. The BiCGstab(2) requires from 1000 to 5000 iterations to converge. After the steady state is computed, computation of the leading eigenvalue was performed using module EB13 of the HSL library. These calculations required about 50,000 Arnoldi iterations for the $100^3$ grid, and about 200,000 iterations for the $256^3$ grid. Note, that such a large amount of the Krylov subspace iterations becomes possible because of the above fast calculation of a next Krylov vector. It should be mentioned also that the EB13 module does not allow one to define explicitly the convergence criterion and numerical tolerance. Possibly, the total number of Arnoldi iterations can be reduced if these parameters are relaxed accordingly to the computational requirements.

Unfortunately, none of the papers where the steady states and the eigenvalues were computed directly, e.g., [14,16-18], did report consumed computational times, so that a direct comparison is impossible. The characteristic times for computation of the steady states were 400 – 600 sec for the $100^3$ grid and 5 – 10 hours for the $256^3$ grid. Calculation of the eigenvalues consumed 1 – 3 hours for the $100^3$ grid and 100 – 200 hours for the $256^3$ grid. Note that in the above cited studies the eigenvalues were calculated on grids containing less than $100^3$ nodes, so that the possibility to perform such calculation on the $256^3$ grids can be considered an achievement, in spite of very long computational times needed.

## 4. Flow visualization technique

To describe our method of visualization of 3D incompressible flows we start from a seemingly simple question of how to compare steady and stable flows calculated for two- and three-dimensional lid-driven cavities at the same Reynolds number. The 2D flow is easily visualized by its streamlines. At the same time, three-dimensional streamlines, as well as trajectories, can be very complicated and cannot be used for a comparison. The obvious and simplest way of plotting the velocity vector field in the 3D cavity midplane will not yield a good comparison for the following reason. Owing to the reflection symmetry, the spanwise velocity component $v$ vanishes in the midplane, however, its spanwise derivative does not. This means that the two-dimensional divergence $\partial u/\partial x + \partial v/\partial y \neq 0$ inside the midplane. Therefore, comparing the vector plots, one would compare a divergence-free 2D vector field with a part of the 3D flow having a non-zero 2D divergence. Clearly, such a comparison cannot be convincing. For a better comparison one can extract the divergence-free part from the vector $(u, 0, w)$ obtained by assigning the spanwise velocity component $v$ of the 3D steady flow to zero. The divergence-free part of this vector, satisfying the boundary conditions (3) and (4), can be obtained by the above proposed SIMPLE-like iteration procedure, as described in [25]. The result is illustrated in Fig. 1, where the divergence-free projections onto the symmetry midplane are compared with the two-dimensional streamlines for $Re = 10$, 100, and 1000. We observe that the isoline patterns are similar at $Re = 10$. At larger Reynolds numbers, owing to the viscous friction at the spanwise walls, the 3D flow intensity in the midplane becomes weaker compared to the 2D one, while the shape of the isolines remains similar. The gray shadowed areas show where in the flow the Bayly criterion for the centrifugal instability [28] is negative. More details on this are given below together with the discussion of three-dimensional instability.

To generalize the example of Fig. 1, we calculate divergence-free projections of the velocity field onto three sets of coordinate planes, (x,y), (y,z), and (x,z). Namely, we compute three projections $v_1$, $v_2$, $v_3$ of the velocity field $v$ onto subspaces formed by divergence-free velocity fields having only two non-zero components. Consider for example coordinate planes (x,z), a subspace formed by all vectors $a$, such that $a = [a_x, 0, a_z]$, and $\nabla \cdot a = \partial a_x/\partial x + \partial a_z/\partial z = 0$. Denote the projection of the velocity field onto this subspace as $v_1$. Similar projections onto the subspaces defined in planes (y,z) and (x,y) are $v_2$ and $v_3$, respectively. Thus, we obtain three vector fields $v_1$, $v_2$, $v_3$, such that each of them has only two non-zero components, and each component is a three-dimensional scalar function. Namely,

$$v_1 = \begin{bmatrix} u_1(x,y,z) \\ 0 \\ w_1(x,y,z) \end{bmatrix}, \quad v_2 = \begin{bmatrix} 0 \\ v_2(x,y,z) \\ w_2(x,y,z) \end{bmatrix}, \quad v_3 = \begin{bmatrix} u_3(x,y,z) \\ v_3(x,y,z) \\ 0 \end{bmatrix}. \tag{11}$$

The two-dimensional divergence of each vector field vanishes:

$$div(v_1) = div_{(x,z)}(v_1) = \frac{\partial u_1}{\partial x} + \frac{\partial w_1}{\partial z} = 0 \tag{12}$$

$$div(v_2) = div_{(y,z)}(v_2) = \frac{\partial v_2}{\partial y} + \frac{\partial w_2}{\partial z} = 0 \tag{13}$$

$$div(v_3) = div_{(x,y)}(v_3) = \frac{\partial u_3}{\partial x} + \frac{\partial v_3}{\partial y} = 0. \tag{14}$$

To ensure that all the three projections are divergence-free and satisfy the boundary conditions we calculate them using a 2D version of the above SIMPLE-like procedure [25]. Then we define a vector potential for each of the vectors, so that each vector potential has only one non-zero component

$$v_1 = rot(\Psi_1); \quad \Psi_1 = \left(0, \Psi^{(y)}(x,y,z), 0\right) \tag{15}$$

$$v_2 = rot(\Psi_2); \quad \Psi_2 = \left(\Psi^{(x)}(x,y,z), 0, 0\right) \tag{16}$$

$$v_3 = rot(\Psi_3); \quad \Psi_3 = \left(0, 0, \Psi^{(z)}(x,y,z)\right) \tag{17}$$

Apparently, the three-dimensional function $\Psi^{(y)}(x,y,z)$ coincides with the stream function of the vector $v_1$ in each plane $y = const$. Therefore, $\Psi^{(y)}$ can be interpreted as an extended two-dimensional stream function, and the same can be said about $\Psi^{(x)}$ and $\Psi^{(z)}$. These potentials are not related, however, to three-dimensional streamlines. As a result, the fields $v_1, v_2$ and $v_3$ are tangent to the corresponding vector potential isosurfaces, and can be interpreted as divergence-free projections of the velocity field onto the coordinate planes. Arguments for uniqueness of these projections and a way to compute them are given in [24,25]. It should be emphasized that the fields $v_1, v_2$ and $v_3$ are not just geometrical projections $(u_x, 0, u_z)$, $(u_x, u_y, 0)$, and $(0, u_y, u_z)$ of the velocity field $u = (u_x, u_y, u_z)$, being visualized, on the coordinates planes. For example, $v_1$ is defined as an orthogonal projection of the vector $(u_x, 0, u_z)$ onto a space of 2D vectors, which are divergence free in the plane $(x, z)$ and satisfy all the boundary conditions. A difference between these vector fields can be seen from the fact that the streamlines of $\boldsymbol{v_1}, \boldsymbol{v_2}$ and $\boldsymbol{v_3}$ are closed, which is not the case for the geometric projections $(u_x, 0, u_z)$, $(u_x, u_y, 0)$, and $(0, u_y, u_z)$.

Examples of this visualization showing three vector potentials, together with the depicted by arrows divergent-free velocity projections onto the corresponding coordinate planes, are presented in Figs. 2 and 3 for the lid motion parallel to the sidewall ($\alpha = 0$, Fig. 2) and parallel to the vertical diagonal plane ($\alpha = 45°$, Fig. 3). The flows are shown at the Reynolds numbers close to their critical values (see below). The bold arrows in all the frames show the direction of the lid motion. To illustrate complicacy of these 3D flows, the frames (a) in both figures represent trajectories of liquid particles calculated from the computed velocity field. The frames (b)-(d) show isosurfaces of the vector potentials, which are supplied by arrow charts of the projected velocities. It is clearly seen that the projected velocity vectors are tangent to the potential isosurfaces.

The main circulation of the case $\alpha = 0$ is located in the (x,z) planes and is depicted in Fig. 2b. As explained in [24], its middle cross-section is one to be compared with the 2D lid-driven square cavity results. In this cross-section we clearly observe the main circulation that can be described by motion in the (x,z) planes only. The fluid moves along the driven lid and then returns along the cube bottom. A small inverse circulation, similar to those observed in the 2D case [1], is also clearly seen near the corner $x = 1, z = 0$. Two other frames, depicted in Figs. 2c and 2d, show three-dimensional "additions" to the main circulation. These "additions" can be divided into two vortical motions. The first one takes place in the (y,z) planes and is most intensive near the $x = 0$ and x=1 cube walls (Fig. 2c). The second one occurs in the (x,y) planes as is depicted in Fig. 2d.

In Fig. 3 the divergence-free projections are made not onto the initial coordinate planes, but on the planes rotated in 45° around the z-axis. This allows us to clearly see the main circulation created by the lid motion (Fig. 3b) and a small inverse vortex in the lower corner. The three-dimensional addition that consists of the two opposite vortical motions in the orthogonal diagonal planes (Fig. 3c) and another, diagonally-symmetric, vertical motion in the (x,y) planes, located mainly in the upper part of the cavity (Fig. 3b). Two bottom frames of Fig. 1 show that the motion in the diagonal plane is noticeably different from the corresponding 2D streamline pattern, which is a consequence of viscous friction that increases from main diagonal towards the cavity corners and causes stronger 3D effects.

## 5. Results and discussion

*5.1 2D lid driven cavity revisited*

Before reporting and discussing results obtained for the 3D flow, and explaining what we possibly can expect from the grid refinement, we revisit results on the 2D lid driven cavity flow stability [5,43-54]. For computation of 2D flows and analysis of their stability we apply the numerical technique of [42], which cannot be extended to 3D flows due to the strong computer memory restrictions. Some published results, together with those obtained in the present study, are collected in Table 1. We observe a noticeable scatter between reported values of the critical Reynolds number and critical frequency. Furthermore, the results obtained with a rather fine spatial resolution [54], exceeding 500 grid nodes in one spatial direction, agree to within the second decimal place only. The main source of this disagreement, in our opinion, is discontinuities of the velocity boundary conditions at the upper cavity corners. Since different numerical methods apply different smoothing of this condition, the resulting steady state solutions slightly differ. Since the eigenvalue problem associated with the linear stability of the flow is ill-conditioned, the results of the eigenvalues calculation differ even stronger. The latter can be seen also from the numerical experiments with different boundary conditions smoothing reported, e.g., in [43] and discussed in [2].

Applying the second order finite volume method, we assume that all the unknown functions vary linearly inside a finite volume. Therefore, for the 2D and $\alpha = 0$ cases, a finer mesh means a steeper jump of the *x*-velocity from the dimensionless unity at the lid to the zero value on the vertical boundaries. On the other hand, the low-order finite volume and finite difference schemes formulated on 5 or 7 points stencils in the 2D or 3D cases, respectively, do not include the corner points in the discretized equations corresponding to the inner points of the flow region. The reader can be referred, e.g., to the schemes of [31], that are applied also in this study. The exclusion of corner points provides an implicit smoothing, which can be considered as "natural" for the chosen numerical method. Keeping this in mind, we can compare results obtained on gradually refined grids to see that the critical Reynolds number and critical frequency converge and, with a sufficient grid refinement, become grid-independent. The latter is illustrated in Table 2 for the 2D lid-driven cavity flow, for which we observe a clear grid convergence of the results. Moreover, assuming that in spite of the discontinuities, the numerical error is $O(h^2)$, where $h$ is a characteristic grid size, we apply the Richardson extrapolation [32]. The result is shown in the two right columns of Table 2, where the Richardson extrapolation is carried out using the current and previous grid sizes. We observe that starting from $300^2$ grid, the Richardson extrapolation is converged to within at least 3 significant decimal places. This approves the assumption made for the numerical error order, and shows that the asymptotically valid Richardson extrapolation is obtained starting from the $300^2$ grid. Furthermore, our result of the Richardson extrapolation is in a very good agreement with the result of [47], so that the two results can be considered as cross validating each other.

Basing on the above 2D results, we expect that the following 3D calculations will also exhibit grid convergence. Clearly, the grid refinement in the 3D computations is strongly restricted compared to the 2D case, by both available computer memory and affordable CPU time. At the same time, the critical Reynolds number in the 3D case [10-18] is approximately 4 times smaller than that of the 2D case, which allows us to expect a better convergence on coarser grids.

## 5.2 *Critical parameters of the lid-driven flow in a cube*

The main quantitative results of the present study are reported in Table 3 and presented graphically in Fig. 4. For the two cases considered we report the critical Reynolds numbers and critical frequencies computed on gradually refined grids. The Richardson extrapolation calculated using the current and previous coarser grid value is shown in bold and italic. We observe that the critical parameters converge monotonically, and we can expect that the Richardson extrapolation yields an improved result, as is shown by dash lines in Fig. 4. It is quite unexpected, however, that two finest results of the Richardson extrapolation coincide within at least 5 significant digits of both the critical Reynolds number and the critical frequency. The converged values are $Re_{cr}(\alpha = 0°) \approx 1919.4$, $\omega_{cr}(\alpha = 0°) = 0.58607$; $Re_{cr}(\alpha = 45°) \approx 2289.3$, $\omega_{cr}(\alpha = 45°) = 0.24931$. It is stressed that results for the $\alpha = 0°$ case are in excellent agreement with those of [14].

It should be mentioned that obtaining above monotonically converging results required quite tough convergence criteria. Thus, for calculation of steady states we had to demand convergence up to the $6^{th}$ decimal place pointwise for all the unknowns. Relaxing this criterion to the $5^{th}$ place altered some of results, which resulted either in a non-monotonic sequence for $\alpha = 0°$, or, for $\alpha = 45°$, in convergence of the Richardson extrapolation within 3-4 significant digits. The eigenvalue solvers used have built-in convergence criteria, iterate until machine convergence and cannot be altered.

Comparison of the critical parameters computed with the same spatial discretization, but using the straightforward time integration [10,22] shows the following. In the $\alpha = 0$ case [10] the calculated critical Reynolds number and critical frequency were 1914, and 0.575, respectively (see also Table 4). While the value of the critical Reynolds number is very close to one obtained here, the critical frequency differs already in the second decimal place. The reason for this can be the subcritical type of this bifurcation, as was argued in [10,14,15]. If the bifurcation is subcritical, the frequency value 0.575 is not expected to be close to the imaginary part of the leading eigenvalue. In this case, the flow transforms abruptly from an unstable steady state to a finite amplitude oscillatory state. In the $\alpha = 45°$ case the result of [22] is $Re_{cr} = 2320, \omega_{cr} = 0.249$. Here, the critical frequency fully agrees with the present results, while the critical Reynolds number is slightly larger. This can be the case of supercritical perturbation, for which the imaginary part of the leading eigenvalue does represent the oscillation frequency, while finite-amplitude oscillation can be observed at the Reynolds number necessarily larger than the critical one.

Table 4 summarizes all published results for the $\alpha = 0$ case. We observe quite a good agreement between all the results, except one of [15] obtained using the large eddy simulation. In this case the disagreement can be a consequence of alteration of the momentum equation by the large eddy model. All other results are not contradictive in the value of the critical Reynolds number. The values of the critical frequencies, reported in [14,16,18], are close to one reported here. Note, that all the three studies report an unstable limit cycle in the slightly supercritical regime. It is worth to mention that the cyclic-fold bifurcation between the upper and lower branches of the subcritical limit cycle was calculated in [18] to be located at $Re \approx 1913.5$, which is close to the value 1914, which is the Richardson extrapolation of the Reynolds numbers corresponding to the appearance of subcritical oscillations at different grids in the time-dependent calculations of [10]. At the same time the three studies [10-12] report qualitatively different supercritical flow states. This discrepancy can result from multiplicity of oscillatory states, from an insufficient accuracy, or from different smoothing of the corner singularities, as discussed in [2]. The unstable limit cycle and the consequent Neimark-Sacker bifurcation that lead to a very complicated temporal behavior were studied in [18].

### 5.3  Patterns of the most unstable perturbations

Several scalar velocity-dependent functions were used in [2,14,18] to visualize three-dimensional oscillatory flow pattern. Here we study patterns of the most unstable perturbation (the leading eigenvector), which show where the flow field is subject to a stronger perturbation, and allow us to make additional conclusions about physical reasons that lead to the instability onset. In the case of the supercritical Hopf bifurcation the absolute value of the components of the perturbation velocity field shows how the oscillation amplitude is distributed in the flow region. As mentioned, some previous works [10,14,15,18] concluded that the steady-oscillatory transition in the case $\alpha = 0$ is subcritical. Recent time-dependent computations performed for $\alpha = 45°$ [22] arrived at a similar conclusion. Clearly, within the linear analysis framework, we cannot examine super- or subcritical character of the observed bifurcations. The conclusions made in the above papers prevent us from interpreting the patterns of absolute value of perturbations as oscillation amplitudes. In the following we compare the patterns of the most unstable perturbations with those reported before, with patterns of the oscillation amplitudes reported in the previous studies, the two cases considered in this study, and with the perturbation pattern of 2D lid driven cavity flow. This allows us to establish similarity between present and previous results, as well as between the $\alpha = 0$ and $45°$ cases. Note that the perturbations reported below, being the eigenvectors, are defined to within multiplication by a complex constant. Therefore, only their relative absolute values are meaningful for comparison with an independent result.

The patterns of the most unstable perturbations are shown in Figs 5 and 6 for the cases $\alpha = 0$ and $45°$, respectively. The isosurfaces levels are shown below the frames. The maximal absolute values are given in the figure captions. Figure 5 contains also perturbation patterns of the two-dimensional lid-driven cavity flow. Note that the absolute values of the perturbation patterns are quite similar to the previously reported ones [16,17], as well as to the oscillation amplitude patterns reported in Fig. 11 of [10].

The two-dimensional perturbations of both velocity components contain two local maxima. One is located in the area where the main circulation rises from the bottom, and does exhibit a certain similarity with the corresponding 3D patterns. The second one, and noticeably larger maximum, is located under the left part of the driven lid. Since the second maximum dominates, the 2D instability most probably sets in inside the boundary layer adjacent to the driven lid. Nothing similar, however, is observed in the three-dimensional patterns (Fig. 5). Moreover, the amplitudes of all three components of velocity perturbation have comparable sizes. In case the instability mechanism was the same as in the 2D case, we would expect smaller amplitudes of the spanwise velocity ($v$) perturbation. Taking into account an almost four times smaller critical Reynolds number of the 3D flows compared to 2D ones, we assume that three-dimensional instability is caused by a three-dimensional physical mechanism, in which all the three velocity components take their parts.

The second important observation that can be derived from Fig. 5, is the symmetry of the real and imaginary parts of the velocity perturbations with respect to the midplane $y = 0.5$. This means that the bifurcation does not break the base flow symmetry, which confirms findings of [18] where the first unstable limit cycle was found to be symmetric. This study describes also further transition from the unstable primary limit cycle to a complicated non-symmetric time-dependent flow.

For the case $\alpha = 45°$ only absolute values of velocity perturbations are shown (Fig. 6). To get a more informative picture, and to compare with the $\alpha = 0$ case, we rotate the coordinate system in $45°$ degrees around the $z$-axis, and plot the resulting velocity components $(u + v)/2$ and $(u - v)/2$ (Fig. 6d and 6e). As shown in the figure, these two components are, respectively, the spanwise and streamwise ones, while the vertical component remains unchanged. These patterns are similar to the oscillation amplitude patterns reported in Figs. 6 and 8 of [22]. Comparing these patterns with those of the $\alpha = 0$ case we do not observe similarities. This can be because the physical mechanism that causes transition to time-dependence is different in the two cases.

The real and imaginary parts of the perturbations (not shown in the figure) of the vertical (Fig. 6c) and the streamwise (Fig. 6e) velocity components are antisymmetric with respect to the diagonal plane indicated by the bold arrows. The corresponding images of the spanwise component perturbation are symmetric, while the patterns shown in Fig. 5a and 5b, as expected, have no any specific symmetry. Thus, the resulting supercritical flow is expected to develop with the break of symmetry, as it was reported in [22].

Some more insight can be gained from the perturbation snapshots distributed over the oscillation period, as shown in Figs. 7 and 8 for the cases $\alpha = 0$ and $45°$, respectively. According to the linear stability analysis the time-dependent perturbation is defined as $Real[\boldsymbol{v}(x, y, z)exp(i\omega t)]$, where $\boldsymbol{v}(x, y, z)$ is the complex eigenvector, and $\omega$ is the imaginary part of the leading eigenvalue. Clearly, the choice of $t = 0$ in this definition is arbitrary. These figures are supplied with animations, which help to understand how the perturbations oscillate in time. In the $\alpha = 0$ case (Fig. 8 and Animation 1) we observe that perturbations of the $x$- and

*y*- velocities appear near the cavity bottom and then are advected by the main circulation. They grow in the area when the circulation turns from horizontal to ascending motion. When the motion becomes purely ascending, the perturbations decay and then disappear. This bifurcation behavior was attributed to the centrifugal instability in [14].

In spite of the isosurfaces patterns in the case $\alpha = 45°$ (Fig. 8 and Animation 2) are noticeably different compared to Fig. 7, we observe a similar trend. Namely, the perturbations start to grow near the cavity bottom, and then are advected by the main circulation, which takes place near the diagonal plane. The perturbations grow when the main circulation motion turns from the horizontal to vertical direction, and then, approaching the driving lid, decay.

Several studies, e.g. [16] and references therein, examine contribution of different terms of the momentum equation to the kinetic energy growth, which is derived via the Reynolds-Orr equation. In our earlier paper [55] we gave several reasons why this approach can be irrelevant for explanation of instabilities that occur owing to the Hopf bifurcation. The two reasons to be mentioned are (i) the Reynolds-Orr equation should be modified when the disturbance vector is complex, and (ii) the potential part of each momentum equation term must be excluded before any further analysis is performed. We do not apply this kind of analysis here.

To make reasonable assumptions about possible physical mechanisms that set the instability in, we note first, that in both cases the disturbances growth is the largest in the neighborhood of the symmetry plane, which is the plane y=0.5 for $\alpha = 0$, and the diagonal plane for $\alpha = 45°$. Then, focusing on the symmetry planes, we compare the main circulation patterns (frames (b) of Figs. 2 and 3) with the patterns of the most unstable perturbations (Figs. 5-8 and Animations 1 and 2). As mentioned, in the both cases the disturbances grow in the area where the main circulation makes a turn from the horizontal to vertical motion. This can be a result of the centrifugal instability [14,56,57], which can be expected to develop where the angular velocity is maximal, i.e., in the neighborhood of the symmetry plane. At the same time, in both cases, the flow contains a small counter vortex, so that the instability may appear as a result of interaction of the two vortices.

In the study [22] it was suggested that the flow oscillations appearing in the $\alpha = 45°$ case, are driven by the interaction of two pairs of vortices depicted in Fig. 3d. In this case, the disturbance appearing in the area where all four vortices meet should be advected from the upper part of the cavity downwards. This contradicts our observation (Fig. 8 and Animation 2), so that the present results do not support this suggestion.

To argue in favor of the centrifugal instability, we follow [56,57] and consider the Rayleigh inviscid stability criterion, which is reformulated for a two-dimensional flow using a quantity $\eta = -\partial |\boldsymbol{r} \times \boldsymbol{u}|^2/\partial r$. The centrifugal instability is expected when the angular momentum decreases outwards a circulation, so that $\eta$ becomes positive. In spite of this criterion is mathematically valid only for inviscid flows, and cannot be applied directly to the viscous ones, it yields a good indication of the presence of centrifugal instability mechanism, especially, if the criterion holds far from the no-slip boundaries. To apply this criterion one needs to know the location of the rotational motion center, so that the radius $\boldsymbol{r}$ will be evaluated from there. Apparently, it is not simple for a fully three-dimensional motion. The task is simplified, however, if we recall that the main circulation is defined by the quasi-two-dimensional projection depicted in Figs. 2b and 3b. This projection extracts the incompressible part of the flow moving inside the two-dimensional coordinate planes. In each cross-section, the center is postulated to be located at the point where the corresponding vector potential attains the maximal absolute value. After the center is found, evaluation of the function $-|\boldsymbol{r} \times \boldsymbol{u}|^2$ is straightforward.

Figures 9a and 9b show the function $-|\boldsymbol{r} \times \boldsymbol{u}|^2$ for both cases considered, in the symmetry plane where the main lid-induced circulation is the most intensive. This circulation is depicted in the midplanes of Figs. 2b and 3b and below we call it the main circulation. Note that we are interested in the region where the function decays with the increasing radius. In the case $\alpha = 0$ (Fig. 9a), we observe that the parameter $\eta$ becomes positive in the region where the main circulation descends and then turns to the horizontal motion along the *x*-axis. Small, but positive values of $\eta$ are observed along the main circulation until it approaches the upper corner. The regions relevant to the Rayleigh criterion are only those containing the turning flow, i.e., areas near the two lower corners. In the $\alpha = 45°$ case (Fig. 9b) we find areas of the positive $\eta$ all the way along the descending part of the main circulation, and even larger $\eta$ in the opposite lower corner when the flow turns to the vertical direction. This can trigger disturbance growth in both regions leading to the complicated patterns of Fig. 6. Thus, in both cases the instability can be triggered by the centrifugal mechanism. At the same time, owing to the distribution of $\eta$, we expected that location of the most perturbed regions with respect to the main circulation is different in the two cases considered. Apparently, such instability should result in vortices located near the flow symmetry plane and altering all the three velocity components.

Another criterion for centrifugal instability proposed by Bayly [28] stays that inviscid 2D flow with closed streamlines is unstable if magnitude of the circulation calculated along a streamline decreases outwards in some region of the flow. Being applied for a viscous flow, it also yields only heuristic indication of possibility of the centrifugal instability mechanism. Considering 3D flows, we again apply it to the circulation in the midplane, which is extracted from the corresponding vector potential. We use the reformulated Bayly criterion applied in [29], which stays that the two-dimensional flow $\boldsymbol{u}_{2D}$ is centrifugally unstable if somewhere inside the flow region the quantity $B = |\boldsymbol{u}_{2D}|\zeta/R$ is negative. Here $\zeta$ in the vorticity of $\boldsymbol{u}_{2D}$, and $R$ is the local algebraic radius of curvature of a streamline, which is expressed and computed as ($\psi$ is the streamfunction of $\boldsymbol{u}_{2D}$)

$$R = \frac{|\boldsymbol{u}_{2D}|^3 \zeta}{(\nabla\psi) \cdot [(\boldsymbol{u}_{2D} \cdot \nabla)\boldsymbol{u}_{2D}]} . \qquad (18)$$

The quantity $B$ was calculated for the streamlines of 2D flows and isolines of the vector potential shown in Fig. 1. Following the representation of [29], the dark gray regions in Fig. 1 correspond to the negative values of $B$. Since the isolines are similar in the 2D and 3D cases, the dark regions are similar as well. This example show that in a viscous flow the quantity $B$ attains negative values even at small Reynolds numbers (e.g., $Re = 10$), at which stability of the flow is doubtless. Isolines of the quantity $B$ at the critical Reynolds numbers are shown in Fig. 9(c,d). Clearly, very large magnitudes of $B$ near the corner discontinuities are due to very large derivatives there, and are irrelevant for the discussion. We observe negative values of $B$ also far from the discontinuities, in the lower part of the flow region. An interesting observation is that location of minima of $B$ is very close to the regions where the Rayleigh criterion holds, so that both inviscid criteria indicate on the possibility of centrifugal instability in the same locations of the base flow.

Recalling the classical centrifugal instability of the rotating Taylor-Couette flow we expect that in the considered case it appears as a pair of vortices rotating in the planes orthogonal to the plane of main circulation. Since radius of the most intensive local rotational motion is located into diagonal plane of the main circulation, and is directed from the cavity center to the lower corner(s), we expect appearance of the vortices in the planes orthogonal to the radius and the above diagonal plane.

Using the above visualization procedure, we can depict the most unstable perturbation in a way that will support or deny the above assumption. To do this, we rotate the *x*- and *z*- axes by the 45° angle, defining new coordinates as

$$x' = (x+z)/2 , \; z' = (x-z)/2, \qquad (18)$$

thus making new axes to be parallel to the diagonal planes. The visualization of a perturbation snapshot for the $\alpha = 0$ is shown in Fig. 10. The green lines in Figs. 10b and 10c indicate the planes, on which orthogonal divergence-free projections are done and where the projected velocity vectors are shown. In Fig. 10a both vector and the base flow streamlines are shown in the midplane.

The frames (b) and (c) of Fig. 10 yield the answer on the posed question: pairs of counter rotating vortices located in the described above planes orthogonal to the radius and the diagonal plane of main circulation. Motion of this vortices takes place around the isosurfaces of $\Psi^{(x')}$ and $\Psi^{(z')}$ This is exactly as expected for the centrifugal instability. The remaining frame (a) shows additional vortices that are shifted from the corners and rotate in the main circulation planes. These additional perturbation motions can result from advection of the centrifugal instability induced vortices. Also, they can relate to another destabilizing mechanism, yet to be described. All the three vector potentials have comparable maximal values reported in the figure caption. Observation of the time evolution of these potentials (Animation 3) shows that the potential $\Psi^{(y')}$ (Fig. 10a) starts to grow in the lower part of the flow, and then is advected by the main circulation, until the corresponding fluid motion is dissipated by the base flow. As it is seen form the supplied animation, velocity projections onto two other planes, $\Psi^{(x')}$ and $\Psi^{(z')}$, (Fig. 10b,c) are not noticeably advected by the main circulation, but mainly grow or decay in their amplitudes, as one would expect for the centrifugal instability.

To produce a similar visualization for the $\alpha = 45°$ case we rotate the coordinates around the z-axis, so that the x-axis becomes parallel to the lid motion direction. Then we rotate again by the angle $\theta = -arccos\left(\sqrt{2/3}\right)$ around the y-axis, so that the resulting x-axis is directed along the diagonal of the diagonal cross-section of the cube. This results in the following coordinate transformation

$$x' = \frac{1}{\sqrt{2}}(x+y)\cos\theta - z\sin\theta, \quad y' = \frac{1}{2}(y-x), \quad z' = \frac{1}{\sqrt{2}}(x+y)\sin\theta - z\cos\theta \quad (19)$$

Result of this visualization is shown in Fig. 11. In this figure the main circulation in shown in the diagonal cross-section plane by the brown potential isolines. Planes, on which the velocity vectors are projected, are shown by green lines in the frames (a) and are shadowed in the frames (b) and (c). Time phase along period of the oscillations for plotting of the snapshots is chosen to show their most characteristic features. The perturbation time evolution can be seen more clearly in Animation 4. In this case the perturbations $\Psi^{(x')}$ and $\Psi^{(z')}$ (Fig. 11b,c) does not appear as pairs of vortices. We observe single vortices, whose directions change during the oscillations period: blue and red colors correspond to the opposite directions of the vortices rotation. The potential $\Psi^{(y')}$ isosurfaces (Fig. 11a) exhibit a pair of counter rotating vortices. The fluid motion in these vortices is parallel to the main diagonal plain, so that they are not orthogonal to the local rotational motion, and therefore cannot result from the centrifugal instability. The counter rotation of these vortices indicates on the symmetry-breaking. Thus, we confirm the observation of [31] where the symmetry-breaking instability of this flow was modeled by a straightforward time integration.

## 6 Conclusions

This study presents a new method for calculation of Krylov vectors for the Krylov-subspace-based iteration methods applied to various problems of incompressible fluid dynamics. The method is applied to the well-known benchmark problem of lid driven flow in a cubical cavity and results in the grid convergent values of the critical Reynolds number and critical frequency calculated for two different directions of the lid motion. These values are calculated via a comprehensive linear stability analysis approach that included the direct computation of steady state flows followed by computation of the leading eigenvalue/eigenvector of the linearized stability problem.

The Newton iteration based steady state solver involved Krylov-subspace iteration methods, BiCGstab(2) and restarted GMRES, for solution of the linear algebraic equations required at each iteration. The Arnoldi eigenvalue solver is by itself a variation of the Krylov-subspace iteration [27]. The Krylov vectors were calculated using the SIMPLE-like procedure [26]. The proposed procedure for calculation of the Krylov vectors does not depend on a problem or on a numerical discretization, so that its applications can be rather wide. Furthermore, the Arnoldi iteration can be carried out in the mode that allows for the direct calculation of the dominant eigenvalue / eigenvector. The latter allowed us to complete the 3D computations on the series of gradually refined grids consisting of $100^3$ to $256^3$ finite volumes.

Basing on the Richardson extrapolation to the 2D lid-driven flow, we applied it also to our 3D results. In both cases the Richardson extrapolated critical values are in a good agreement with the previously published results obtained by the straightforward integration in time or eigenvalues computation.

By comparison of the linear stability findings and the time-dependent non-linear results, we confirmed the subcritical character of the bifurcation in the $\alpha = 0$ case, as was suggested in [10,14,15], and was studied in detail in [18]. We did not gain, however, any additional arguments to support the suggestion of [22] about the subcritical character of the bifurcation in the $\alpha = 45°$ case. Contrarily, a good agreement of the critical frequencies and a slightly smaller value of the critical Reynolds number obtained in this work, indicate on the supercritical type of this bifurcation.

It was found that for $\alpha = 0$ the perturbation remains symmetric with respect to the $y = 0.5$ plane. According to the results of [18] the corresponding limit cycle is unstable, so that resulting oscillatory flow is non-symmetric, as was reported also in [10]. For $\alpha = 45°$ the most unstable disturbance is non-symmetric, so that the instability necessarily results in an non-symmetric oscillatory flow, as it was observed in [22].

The flow and most unstable disturbance patterns were visualized by a novel approach proposed in [24,25]. Among a better insight in the three-dimensional structure of the velocity and disturbances fields, we used this visualization method for a consistent evaluation of the Rayleigh and Bayly instability criteria and graphical illustration of additional vortices induced in the flow by instability. Basing on the above, we offered some arguments regarding the physical mechanisms that make the flow unstable. In particular, assuming that the Rayleigh and Bayly inviscid criteria can be meaningful when is evaluated for the main circulation, we use formulation of [57] for its evaluation. The criteria hold in both cases considered, indicating that the instability can be driven by the centrifugal mechanism. To verify this, we visualized the most unstable perturbation in a way that allowed us to see whether the instability appears as a pair of counter rotating vortices in the planes orthogonal to the rotational motion of the main circulation, as it happens, e.g., in primary instability of the Taylor-Couette flow. This visualization allowed us to confirm the centrifugal mechanism for the $\alpha = 0$ case and to deny it for the $\alpha = 45°$ case. Similarly, the assumption of [22] that linked the instability to interaction of the counter rotating vortices is not supported by our visualization plots.

# References

1. Shankar, P.N., Deshpande, M.D.: Fluid mechanics in the driven cavity. Ann. Rev. Fluid Mech. 32, 93-136 (2000)
2. Kuhlmann, H.C., Romano F.: The lid-driven cavity. In: Computational Modelling of Bifurcations and Instabilities in Fluid Dynamics (ed. A. Gelfgat), Springer, 2018, 233-310.
3. Deshmuck R., McNamara, J.J., Liang, Z., Kolter J.Z.: Abhijit G.: Model order reduction using sparse coding exemplified for the lid-driven cavity. J. Fluid Mech., 808, 189-223 (2016)
4. Kalita, J.C., Gogoi, B.B.: A biharmonic approach for the global stability analysis of 2D incompressible viscous flows. Applied Mathematical Modelling, 40, 6831-6849 (2016)

5. Nuriev, A.N., Egorov, A.G., Zaitseva O.N.: Bifurcation analysis of steady-state flows in the lid-driven cavity. Fluid Dyn. Res. 48, 061405 (2016)
6. Babu, V., Korpela, S.A.: Numerical solution of the incompressible, three-dimensional Navier-Stokes equations. Computers & Fluids, 23, 675-691 (1994)
7. Albensoeder, S., Kuhlmann, H.C.: Accurate three-dimensional lid-driven cavity flow. J. Comput. Phys., 206, 536-558 (2006)
8. Liberzon, A., Feldman, Y., Gelfgat, A.Y.: Experimental observation of the steady – oscillatory transition in a cubic lid-driven cavity. Phys Fluids, 23, 084106 (2011)
9. Feldman, Y., Gelfgat, A.Y.: On pressure-velocity coupled time-integration of incompressible Navier-Stokes equations using direct inversion of Stokes operator or accelerated multigrid technique. Computers and Structures, 87, 710-720 (2009)
10. Feldman, Y., Gelfgat, A.Y.: Oscillatory instability of a 3D lid-driven flow in a cube. Phys. Fluids 22, 093602 (2010)
11. Hammami, F., Ben-Cheikh, N., Campo, A., Ben-Beya, B., Lili, T.: Prediction of unsteady states in Lid-driven cavities filled with an incompressible viscous fluid. Int. J. Modern Phys. C, 23, 1250030 (2012)
12. Mynam, M., Pathak, A.D.: Lattice Boltzmann simulation of steady and oscillatory flows in lid-driven cubic cavity. Int. J. Modern Phys. C, 24, 1350005 (2013)
13. Chang, H.W., Hong, P.Y., Lin, L.S., Lin, C.A.: Simulations of flow instability in three dimensional deep cavities with multi relaxation time lattice Boltzmann method on graphic processing units. Computers & Fluids, 88, 866-871 (2013)
14. Kuhlmann, H.C., Albensoeder, S.: Stability of the steady three-dimensional lid-driven flow in a cube and the supercritical flow dynamics, Phys. Fluids 26: 024104 (2014)
15. Anupindi, K., Lai, W., Frankel, S.: Characterization of oscillatory instability in lid driven cavity flows using lattice Boltzmann method, Computers & Fluids 92, 7-21 (2014)
16. Loiseau, J.C., Robinet, J.C., Leriche, E.: Intermittency and transition to chaos in the cubical lid-driven cavity flow. Fluid Dyn. Res. 48, 061421 (2016)
17. Gómez, F., Gómez, R., Theofilis, V.: On three-dimensional global linear instability analysis of flows with standard aerodynamics codes. Aerospace Science and Technology, 32, 223-234 (2014)
18. Lopez, J.M., Welfert, B.D, Wu, K., Yalim, J.: Transitions to complex dynamics in the cubic lid-driven cavity, Phys. Rev. Fluids, 2, 074401 (2017).
19. Povitsky, A.: High-incidence 3-D lid-driven cavity flow. AIAA Paper, 2847 (2001)
20. Povitsky, A.: Three-dimensional flow in cavity at yaw. Nonlinear Anal. Theory Methods Appl. 63, e1573–e1584 (2005)
21. Feldman, Y., Gelfgat, A.Y.: From multi- to single-grid CFD on massively parallel computers: numerical experiments on lid-driven flow in a cube using pressure–velocity coupled formulation. Computers & Fluids 46, 218–223 (2011)
22. Feldman, Y.: Theoretical analysis of three-dimensional bifurcated flow inside a diagonally lid-driven cavity. Theor. Comput. Fluid Dyn. 29, 245–261 (2015)
23. Gulberg, Y., Feldman, Y.: On laminar natural convection inside multi-layered spherical shells, Int. J. Heat Mass Transfer, 91, 908-921 (2015)
24. Gelfgat, A.Y.: Visualization of three-dimensional incompressible flows by quasi-two-dimensional divergence-free projections, Computers & Fluids 97, 143-155 (2014)

25. Gelfgat, A.Y.: Visualization of three-dimensional incompressible flows by quasi-two-dimensional divergence-free projections in arbitrary flow regions, Theor. Comput. Fluid Dyn. 30, 339-348 (2016).
26. Patankar, S.V.: Numerical Heat Transfer and Fluid Flow. Taylor & Francis, 1980.
27. van der Vorst H. Iterative Krylov Methods for Large Linear Systems. Cambridge Univ. Press, 2003.
28. Bayly B.J.: Three-dimensional centrifugal-type instabilities in inviscid two-dimensional flows. Phys. Fluids, 31, 56-64 (1988).
29. Lanzerstorfer D., Kuhlmann H.C.: Global stability of the two-dimensional flow over a backward-facing step. J. Fluid Mech., 693, 1-27 (2012).
30. Albensoeder S., Kuhlmann H.C., Rath H.J.: Three-dimensional centrifugal-flow instabilities in the lid-driven-cavity problem. Phys. Fluids, 13, 121-135 (2001).
31. Feldman, Y.: Direct numerical simulation of transitions and supercritical regimes in confined three-dimensional recirculating flows, PhD Thesis, Tel-Aviv University, 2010.
32. Roache P.J.: Perspective: a method for uniform reporting of grid refinement studies. J. Fluids Eng., 116, 405-413 (1994).
33. Sorensen, D.C.: Implicit application of polynomial filters in a k-step Arnoldi method. *SIAM J. Matrix Analysis and Applications* 13, 357-385 (1992)
34. Scott, J.A.: An Arnoldi code for computing selected eigenvalues of sparse real unsymmetric matrices. ACM Trans. Math. Software 21, 432-475 (1995)
35. Edwards, W.S., Tuckerman, L.S., Friesner, R.A., Sorensen, D.C.: Krylov methods for the incompressible Navier-Stokes equations. J. Comput. Phys. 110, 82-102 (1994)
36. Tuckerman, L.S., Barkley, D.: Bifurcation analysis for time-steppers, in "Numerical Methods for Bifurcation Problems and Large-Scale Dynamical Systems", (eds. K. Doedel & L. Tuckerman). IMA Volumes in Mathematics and Its Applications, 119, Springer, New York, 453-466 (2000)
37. Tuckerman, L.S., Bertagnolio, F., Daube, O., Le Quéré, P., Barkley, D.: Stokes preconditioning for the inverse Arnoldi method, in "Continuation Methods for Fluid Dynamics" (Notes on Numerical Fluid Dynamics, 74), eds. D. Henry and A. Bergeon, Vieweg, Göttingen, 241-255 (2000)
38. Gelfgat, A.Y.: Krylov-subspace-based steady state and stability solvers for incompressible flows: replacing time steppers and generation of initial guess, In: Computational Modelling of Bifurcations and Instabilities in Fluid Dynamics (ed. A. Gelfgat), Springer, 2018 (to appear).
39. Chorin, A.J.: Numerical solution of the Navier-Stokes equations, Mathematics of Computation 22, 745-762 (1968 )
40. Vitoshkin, H., Gelfgat, A.Y.: On direct inverse of Stokes, Helmholtz and Laplacian operators in view of time-stepper-based Newton and Arnoldi solvers in incompressible CFD, Comm. Comput. Phys. 14, 1103-1119 (2013)
41. Lynch, R.E., Rice, J.R., Thomas, D.H.: Direct solution of partial differential equations by tensor product methods, Numer. Math. 6,185-199 (1964)
42. Gelfgat, A.Y.: Stability of convective flows in cavities: Solution of benchmark problems by a low-order finite volume method, Intl. J. Num. Meth. Fluids 53, 485-506 (2007)

43. Gelfgat, A.Y.: Implementation of arbitrary inner product in global Galerkin method for incompressible Navier-Stokes equation, *J. Comput. Phys*. 211, 513-530 (2006)
44. Poliashenko, M., Aidun, C,K.: A direct method for computation of simple bifurcations. J. Comput. Phys. 121, 246-260 (1995)
45. Gervais, J.J., Lemelin, D., Pierre, R.: Some experiments with stability analysis of discrete incompressible flows in the lid-driven cavity. Int. J. Numer. Meth. Fluids 24, 477-492 (1997)
46. Fortin, A., Jardak, M., Gervais, J.J., Pierre, R.: Localization of Hopf bifurcations in fluid flow problems. Int. J. Numer. Meth. Fluids 24, 1185-1210 (1997)
47. Auteri, F., Parolini, N., Quartapelle, L.: Numerical investigations on the stability of singular driven cavity flow. J. Comput. Phys. 183, 1-25 (2002)
48. Peng, Y.F., Shiau, Y.H., and Hwang, R.R.: Transition in a 2-D lid-driven cavity flow. Computers & Fluids 32, 337-352 (2003)
49. Abouhamza, A., Pierre, R.: A neutral stability curve for incompressible flows in a rectangular driven cavity. Math. Comput. Modelling, 38, 141-157 (2003)
50. Cadou, J.M., Potier-Ferry, M., Cochelin B.: A numerical method for the computation of bifurcation points in fluid mechanics. Eur. J. Mech. B/Fluids 25, 234-254 (2006)
51. Sahin, M., Owens, R.G.: A novel fully-implicit finite volume method applied to the lid-driven cavity problem. Part II. Linear stability analysis. Int. J. Numer. Meth. Fluids 42, 79-88 (2003)
52. Boppana, V.B.L., Gajjar, J.S.B.: Global flow instability in a lid-driven cavity. Int. J. Numer. Meth. Fluids 62, 827-853 (2010)
53. Tiesinga, G., Wubs, F.W., Veldman, A.E.P.: Bifurcation analysis of incompressible flow in a driven cavity by the Newton–Picard method. J. Comput. Appl. Math. 140, 751-772 (2002)
54. Kalita, J.C., Gogoi, B.B.: A biharmonic approach for the global stability analysis of 2D incompressible viscous flows. Appl. Math. Modelling 40, 6831-6849 (2016)
55. Gelfgat, A.Y., Molokov, S.: Quasi-two-dimensional convection in a 3D laterally heated box in a strong magnetic field normal to main circulation. Phys. Fluids 23, 034101 (2011)
56. Brès, C.A., Colonius, T.: Three-dimensional instabilities in compressible flow over open cavities. J. Fluid Mech. 599, 309-339 (2008)
57. Barkley, D., Gomes, M.G.M.: Henderson RD. Three-dimensional instability in flow over a backward-facing step. J. Fluid Mech. 473, 167-190 (2002)

**Figure captions**

Figure 1. Streamlines of two-dimensional lid-driven cavity flows (left column) compared with the divergence-free projections of 3D lid-driven flow onto the symmetry midplane (right column). The (positive) streamlines of the main circulation are equally spaced between 0.01 and 0.1 with the step 0.01. The negative streamlines are added only for visualization purposes. Shadowed areas correspond to the negative quantity $B$ of the Bayly criterion.

Figure 2. Visualization of 3D velocity fields corresponding to a slightly subcritical steady state. $Re = 1900, \alpha = 0$. (a) trajectories computed from the calculated velocities. (b) – (d) Divergence-free projections of velocity fields onto the coordinate planes. The projected velocities fields are depicted by vectors. Isosurfaces of the velocity potentials, to which the projected velocities fields are tangent, are shown by colors. The minimal and maximal values of the potentials are $\pm 0.014$, $(-0.063, 0.0042)$ and $\pm 0.0087$ for $\Psi^{(x)}$, $\Psi^{(y)}$, and $\Psi^{(z)}$, respectively. The isosurfaces are plotted for the levels at $\pm 0.005$ for $\Psi^{(x)}$, $\pm 0.004$ for $\Psi^{(z)}$, and $0.0005$ and $-0.04$ for $\Psi^{(y)}$.

Figure 3. Visualization of 3D velocity fields corresponding to slightly subcritical steady states. $Re = 2300, \alpha = 45°$. (a) trajectories computed from the calculated velocities. (b) – (d) Divergence-free projections of velocity fields onto the coordinate planes after the coordinates were rotated in 45°, so that the lid moves along the rotated $y$ axis. The projected velocities fields are depicted by vectors. Isosurfaces of the velocity potentials, to which the projected velocities fields are tangent, are shown by colors. The minimal and maximal values of the potentials are $\pm 0.020$, $(-0.012, 0.092)$ and $\pm 0.018$ for $\Psi^{(x)}$, $\Psi^{(y)}$, and $\Psi^{(z)}$, respectively. The isosurfaces are plotted for the levels at $\pm 0.015$ for $\Psi^{(x)}$, $\pm 0.014$ for $\Psi^{(z)}$, and $0.02$ and $-0.006$ for $\Psi^{(y)}$.

Figure 4. Calculated critical Reynolds numbers and frequencies versus the grid refinement. $N$ is number of grid points in one spatial direction. Dash lines show extrapolation to a zero grid size (Richardson extrapolation).

Figure 5. Isosurfaces of components of the leading velocity eigenvector for the case $\alpha = 0$. Absolute value of the leading 2D disturbance of the $x$- and $z$- velocities is shown as 2D plots in the right hand side of the figure. Maximal absolute values of the perturbations are 0.0085, 0.0059, and 0.0069 for $v_x$, $v_y$ and $v_z$, respectively.

Figure 6. Isosurfaces components of of the leading velocity eigenvector absolute value for the case $\alpha = 45°$. The velocity perturbations are plotted in the coordinates rotated around the $z$-axis, so that the lid moves parallel to the diagonal plane in the direction shown by the bold arrows. The maximal absolute values are 0.0054 for the perturbation of the streamwise velocity (a), 0.003 for the spanwise one (b), and 0.0056 for the perturbation of $v_z$ (c).

Figure 7. Isosurfaces of the leading velocity eigenvector at four equally distanced times along the oscillation period for the case $\alpha = 0$ (Animation 1). The values plotted are $\pm 0.0015$

for each velocity component. The minimal and maximal values are -0.00850 and 0.00568 for $v_x$, ±0.00569 for $v_y$, and -0.00411 and 0.00632 for $v_z$.

Figure 8. Isosurfaces of the leading velocity eigenvector at four equally distanced times along the oscillation period for the case $\alpha = 45°$ (Animation 2). The coordinate system is rotated in 45 around the z-axis. The values plotted are ±0.0015 for each velocity component. The minimal and maximal values are -0.00342 and 0.00427 for $v_x$, -0.00775 and 0.00732 for $v_y$, and ±0.00823 for $v_z$.

Figure 9. Function $-(r \times v)^2$ of the Rayleigh stability criterion (a,b) and the quantity $B$ of the Bayly criterion (c,d) computed at the symmetry midplane for the slightly subcritical flows at (a) $\alpha = 0$ and (b) $\alpha = 45°$.

Figure 10. Visualization of the most unstable perturbation snapshots in axes rotated by Eq. (18), for the $\alpha = 0$ case. Brown streamlines correspond to the vector potential $\Psi^{(y\prime)}$ of the base flow. The isosurfaces and the arrows show the vector potentials and the divergence-free velocity projections calculated for the perturbation snapshots (Animation 3). The isosurfaces are plotted for the levels -0.00023 and 0.00028 for $\Psi^{(y\prime)}$, ±0.0003 for $\Psi^{(x\prime)}$, and ±0.00022 for $\Psi^{(z\prime)}$. The minimal and maximal values of the vector potentials are -0.0630 and 0.00417 for $\Psi^{(y\prime)}$, -0.0142 and 0.01367 for $\Psi^{(x\prime)}$, and ±0.00868 for $\Psi^{(z\prime)}$.

Figure 11. Visualization of the most unstable perturbation snapshots in axes rotated by Eq. (19), for the $\alpha = 45°$ case. Brown streamlines correspond to the vector potential $\Psi^{(y\prime)}$ of the base flow. The isosurfaces and the arrows show the vector potentials and the divergence-free velocity projections calculated for the perturbation snapshots (Animation 4). Gray color indicates planes in which the arrows are plotted. The isosurfaces are plotted for the levels ±2.5×10⁻⁵ for $\Psi^{(y\prime)}$, $\Psi^{(x\prime)}$, and $\Psi^{(z\prime)}$. The minimal and maximal values of the vector potentials are -1.23×10⁻⁴ and 5.42×10⁻⁴ for $\Psi^{(y\prime)}$, -2.15×10⁻⁴ and 2.90×10⁻⁴ for $\Psi^{(x\prime)}$, and -3.94×10⁻⁴ and 3.81×10⁻⁴ for $\Psi^{(z\prime)}$.

Table 1. Critical Reynolds number and critical oscillation frequency computed in different studies for the 2D lid-driven cavity flow.

| Reference | Discretization | $Re_{cr}$ | $\omega_{cr}$ |
|---|---|---|---|
| [44] | $57^2$ finite biquadratic elements | 7763.4 | 2.8634 |
| [45] | Unstructured finite element grid with 27890 degrees of freedom | 8040. | 2.829 |
| [46] | $60^2$ finite biquadratic elements | 8000. | 2.8356 |
| [47] | $160^2$ Chebyshev collocation points | 8018. | 2.8249 |
| [48] | $200^2$ uniform grid | 7704. | 3.707 |
| [49] | 1936 triangle FE | 8004.5 | 2.856 |
| [50] | 13,122 degrees of freedom, FE irregular mesh | 7890 | 2.765 |
| [51] | $257^2$ stretched grid | 8069.76 | 2.8251 |
| [43] | 60×60 global Galerkin basis functions | 7975. | 2.829 |
| [52] | 501×121 FD × collocation grid | 8026.6 | 2.8256 |
| [53] | $128^2$ FV grid | 8375 | 2.76 |
| [54] | $501^2$ FD grid | 8025.9 | 2.81642 |
| [5] | $512^2$ FD grid | 8141 | 2.7909 |
| present | Richardson extrapolation from $550^2$ and $600^2$ FV grids | 8018 | 2.8287 |

Table 2. Convergence of the critical Reynolds number and the critical frequency with the grid refinement. 2D lid-drivem flow in a square cavity. The Richardson extrapolation calculated via the current and the previous coarser grid.

| Grid $N=N_x=N_y$ | $Re_{cr}$ | $\omega_{cr}$ | $Re_{cr}$, Richardson extrapolation | $\omega_{cr}$, Richardson extrapolation |
|---|---|---|---|---|
| 100 | 8264.075 | 2.272823 | | |
| 200 | 8068.25 | 2.80328 | 8002.98 | 2.98010 |
| 300 | 8040.76 | 2.81751 | 8018.77 | 2.82889 |
| 400 | 8031.075 | 2.82219 | 8018.623 | 2.82821 |
| 500 | 8026.55 | 2.82453 | 8018.51 | 2.82869 |
| 600 | 8024.07 | 2.8258 | 8018.43 | 2.82869 |

Table 3. Convergence of the critical Reynolds number and the critical frequency with the grid refinement. The Richardson extrapolation calculated via the current and the previous coarser grid is shown by the bold italic letters.

| Grid | $100^3$ | | $150^3$ | | $200^3$ | | $256^3$ | |
|---|---|---|---|---|---|---|---|---|
| Case | $Re_{cr}$ | $\omega_{cr}$ | $Re_{cr}$ | $\omega_{cr}$ | $Re_{cr}$ | $\omega_{cr}$ | $Re_{cr}$ | $\omega_{cr}$ |
| $\alpha = 0$ | 1963.74 | 0.575370 | 1938.98 | 0.581315 | 1930.40 | 0.583398 | 1926.10 | 0.584438 |
| ***Richardson extr.*** | | | ***1919.16*** | ***0.58607*** | ***1919.37*** | ***0.58607*** | ***1919.37*** | ***0.58607*** |
| $\alpha = 45°$ | 2266.29 | 0.250855 | 2278.39 | 0.250305 | 2283.17 | 0.249868 | 2285.56 | 0.249650 |
| ***Richardson extr.*** | | | ***2288.06*** | ***0.24986*** | ***2289.31*** | ***0.249307*** | ***2289.31*** | ***0.249307*** |

Table 4. Critical Reynolds number and critical oscillation frequency computed in different studies for the 3D lid-driven cavity flow at $\alpha = 0$.

| Source | Discretization | Method | $Re_{cr}$ | $\omega_{cr}$ |
|---|---|---|---|---|
| [10] | $100^3$ to $200^3$ finite volumes and Richardson extrapolation | time integration | 1914 | 0.575 |
| [11] | $48^3$ finite volumes | time integration | 1922 | |
| [12] | $301^3$ | Lattice Boltzmann simulation | $1900 < Re_{cr} < 2000$ | |
| [13] | $128^3$ | Lattice Boltzmann simulation | $1750 < Re_{cr} < 1950$ | |
| [14] | $128^3$ Chebyshev collocation | time integration | 1919.5 | 0.586 |
| [15] | $80^3$ | Large eddy simulation | $2100 < Re_{cr} < 2250$ | 0.012 |
| [16] | 1000 spectral elements of the 6$^{th}$ order | Eigenvalue analysis | $1900 < Re_{cr} < 1930$ | 0.585 |
| [18] | $48^3$ spectral collocation modes | Eigenvalue analysis | 1928.9 | 0.5832 |
| present | $256^3$ finite volume grid | Eigenvalue analysis | 1919.4 | 0.5861 |

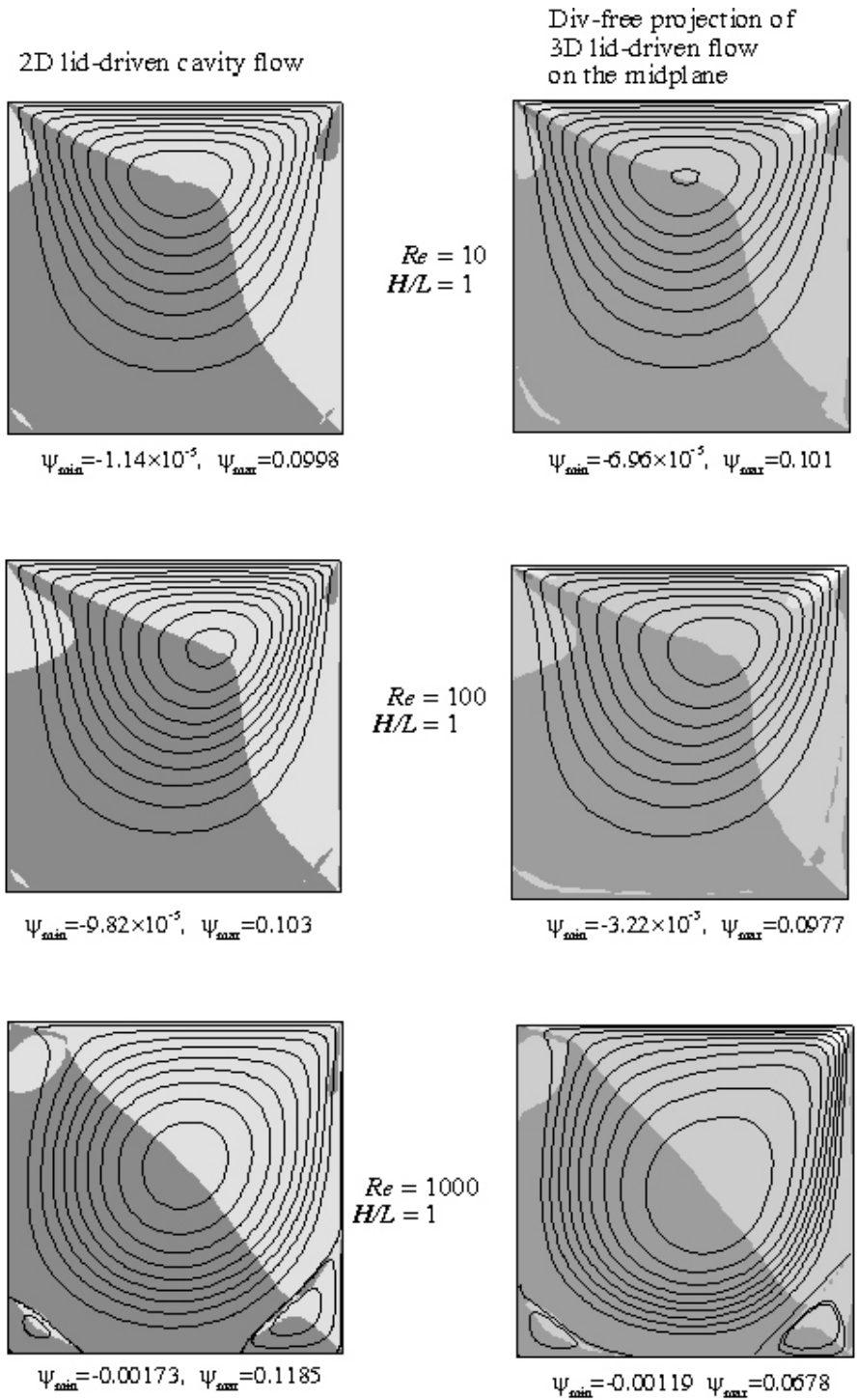

Figure 1. Streamlines of two-dimensional lid-driven cavity flows (left column) compared with the divergence-free projections of 3D lid-driven flow onto the symmetry midplane (right column). The (positive) streamlines of the main circulation are equally spaced between 0.01 and 0.1 with the step 0.01. The negative streamlines are added only for visualization purposes. Shadowed areas correspond to the negative quantity $B$ of the Bayly criterion.

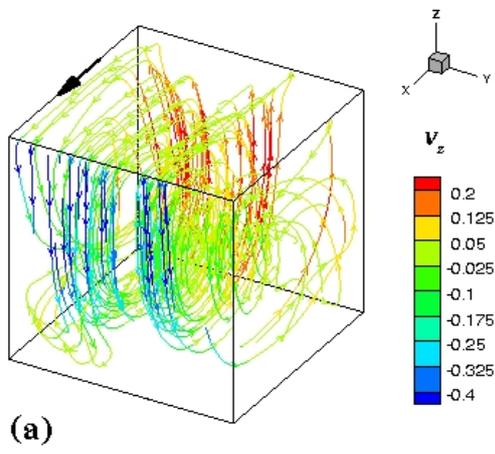
(a)

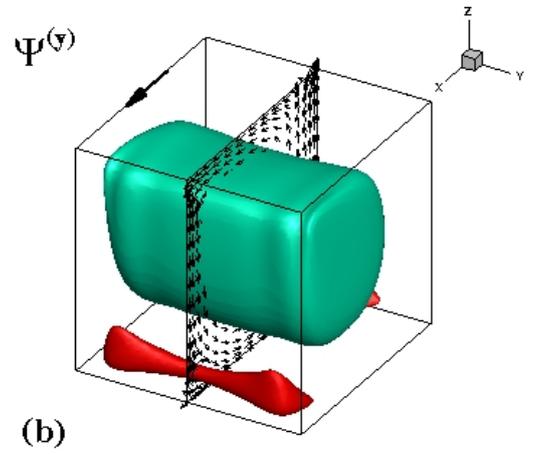
(b)

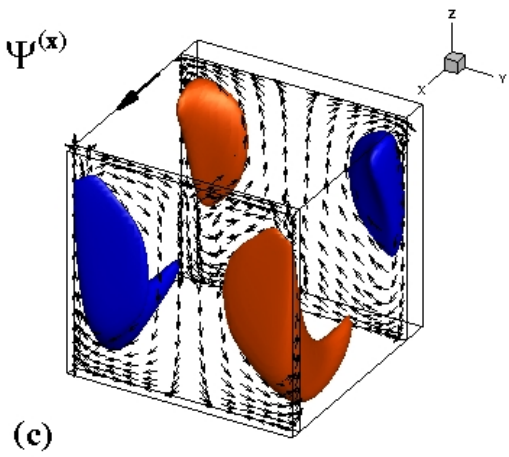
(c)

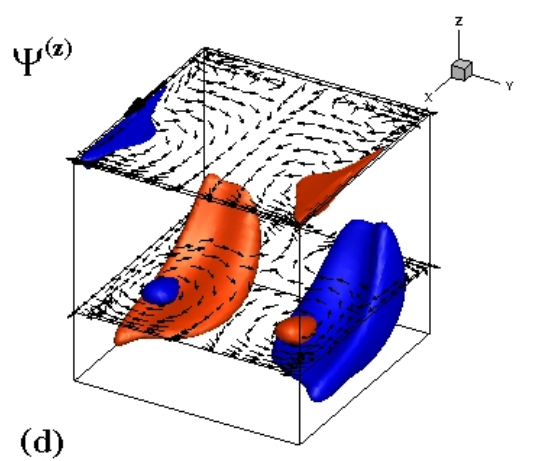
(d)

Figure 2. Visualization of 3D velocity fields corresponding to aslightly subcritical steady state. $Re = 1900, \alpha = 0$. (a) Trajectories computed from the calculated velocities. (b) – (d) Divergence-free projections of velocity fields onto the coordinate planes. The projected velocities fields are depicted by vectors. Isosurfaces of the velocity potentials, to which the projected velocities fields are tangent, are shown by colors. The minimal and maximal values of the potentials are $\pm 0.014$, $(-0.063, 0.0042)$ and $\pm 0.0087$ for $\Psi^{(x)}$, $\Psi^{(y)}$, and $\Psi^{(z)}$, respectively. The isosurfaces are plotted for the levels at $\pm 0.005$ for $\Psi^{(x)}$, $\pm 0.004$ for $\Psi^{(z)}$, and $0.0005$ and $-0.04$ for $\Psi^{(y)}$.

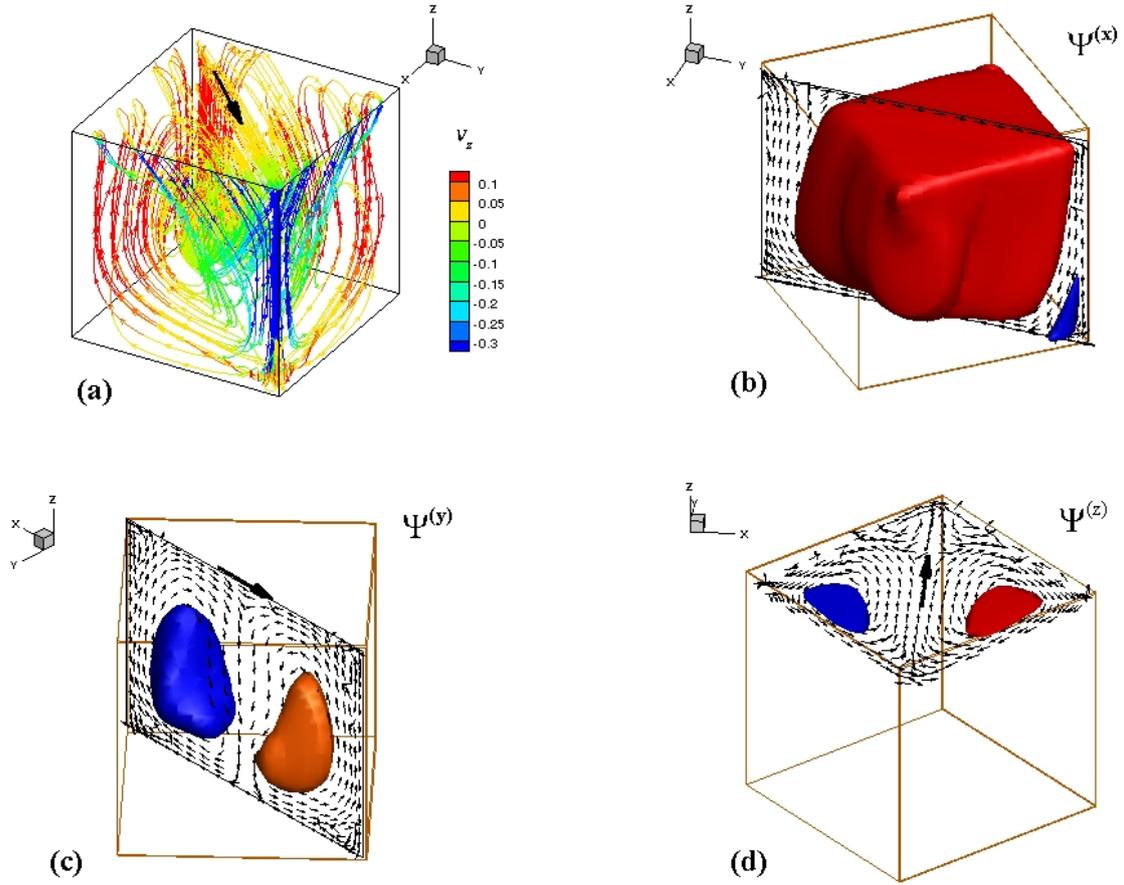

Figure 3. Visualization of 3D velocity fields corresponding to slightly subcritical steady states. $Re = 2300, \alpha = 45°$. (a) trajectories computed from the calculated velocities. (b) – (d) Divergence-free projections of velocity fields onto the coordinate planes after the coordinates were rotated at 45°, so that the lid moves along the rotated $y$ axis. The projected velocities fields are depicted by vectors. Isosurfaces of the velocity potentials, to which the projected velocities fields are tangent, are shown by colors. The minimal and maximal values of the potentials are $\pm 0.020$, $(-0.012, 0.092)$ and $\pm 0.018$ for $\Psi^{(x)}$, $\Psi^{(y)}$, and $\Psi^{(z)}$, respectively. The isosurfaces are plotted for the levels at $\pm 0.015$ for $\Psi^{(x)}$, $\pm 0.014$ for $\Psi^{(z)}$, and 0.02 and $-0.006$ for $\Psi^{(y)}$.

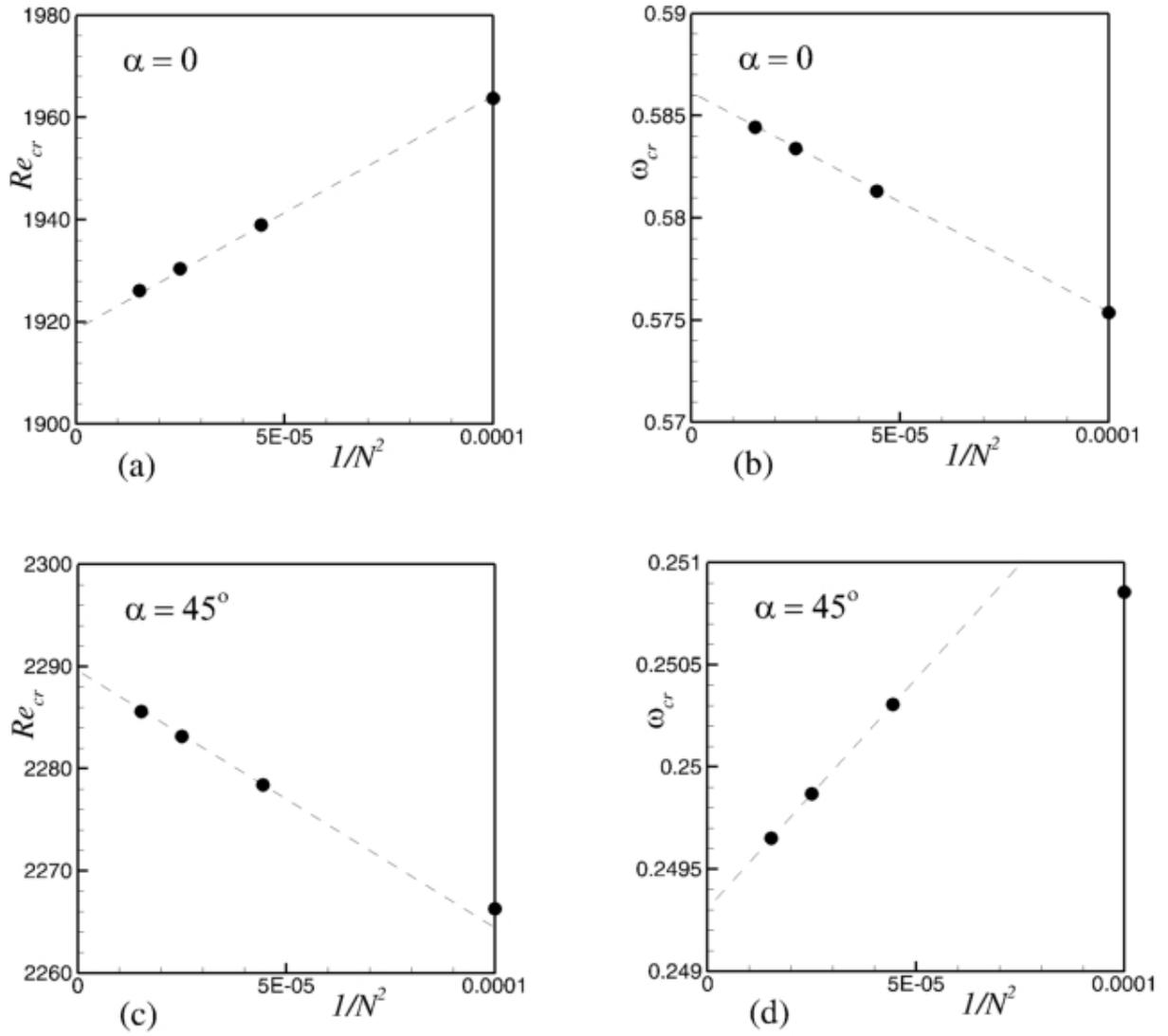

Figure 4. Calculated critical Reynolds numbers and frequencies versus the grid refinement. $N$ is number of grid points in one spatial direction. Dash lines show extrapolation to a zero grid size (Richardson extrapolation).

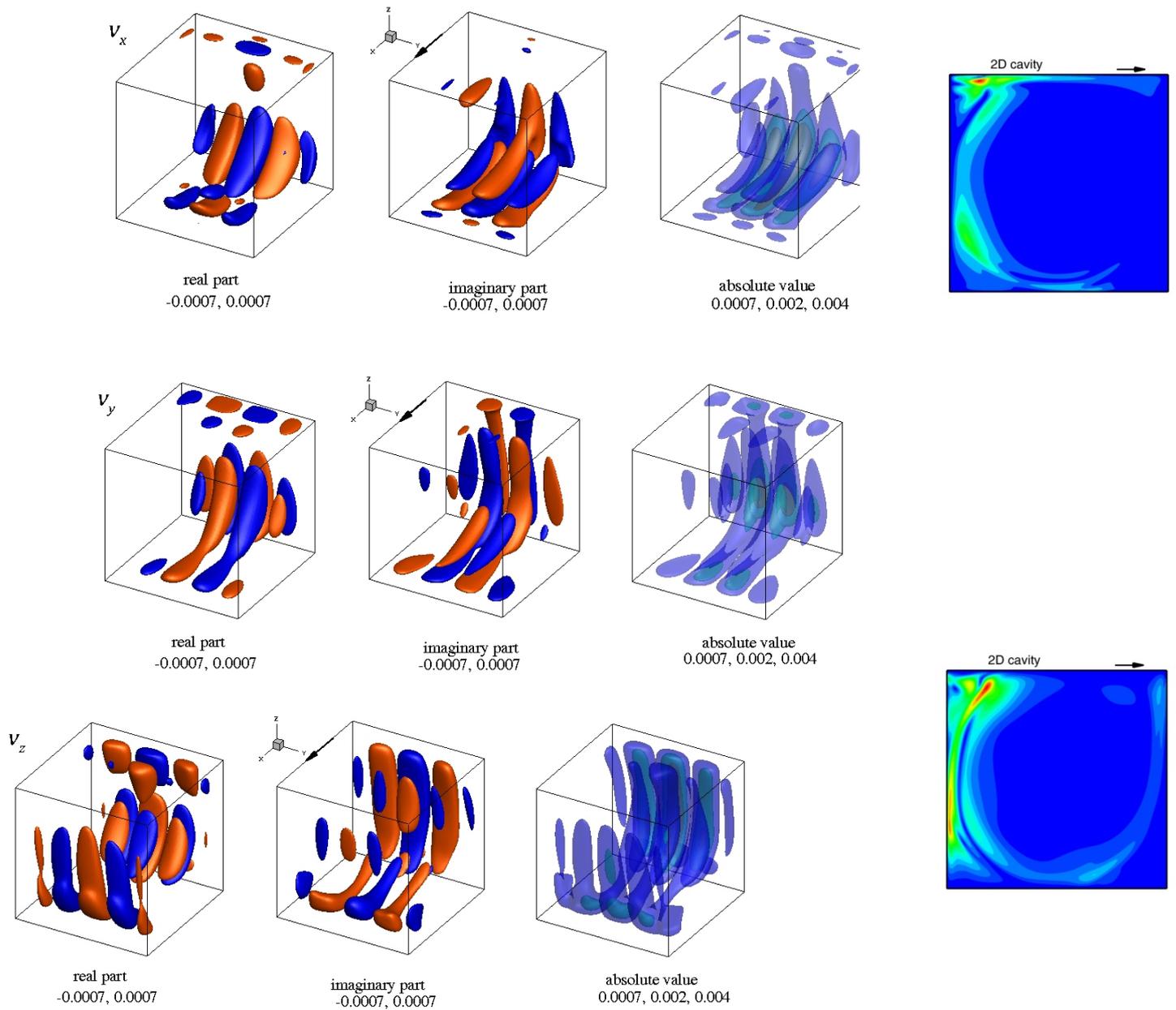

Figure 5. Isosurfaces of components of the leading velocity eigenvector for the case $\alpha = 0$. Absolute value of the leading 2D disturbance of the *x*- and *z*- velocities is shown as 2D plots in the right hand side of the figure. Maximal absolute values of the perturbations are 0.0085, 0.0059, and 0.0069 for $v_x$, $v_y$ and $v_z$, respectively.

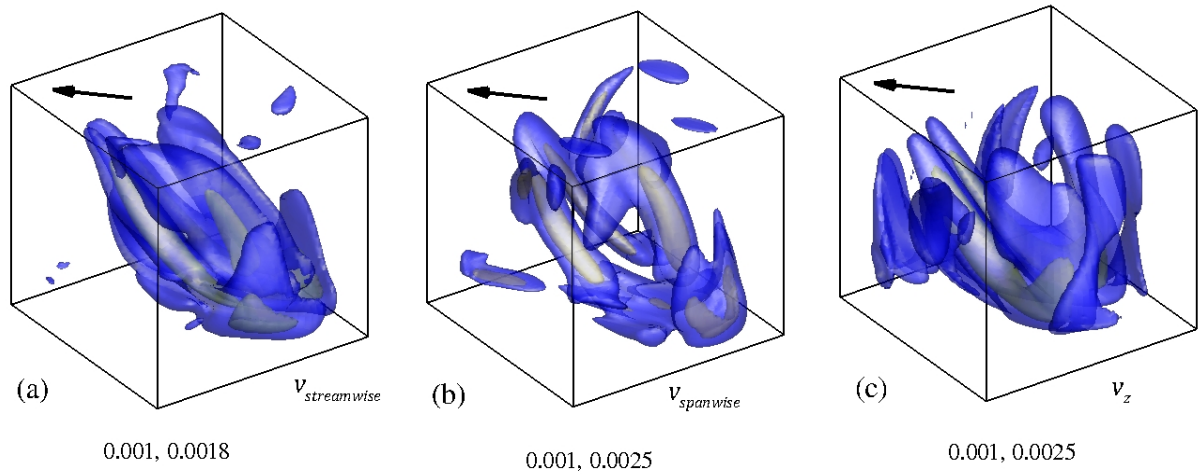

Figure 6. Isosurfaces of components of the leading velocity eigenvector absolute value for the case $\alpha = 45°$. The velocity perturbations are plotted in the coordinates rotated around the $z$-axis, so that the lid moves parallel to the diagonal plane in the direction shown by the bold arrows. The maximal absolute values are 0.0054 for the perturbation of the streamwise velocity (a), 0.003 for the spanwise one (b), and 0.0056 for the perturbation of $v_z$ (c).

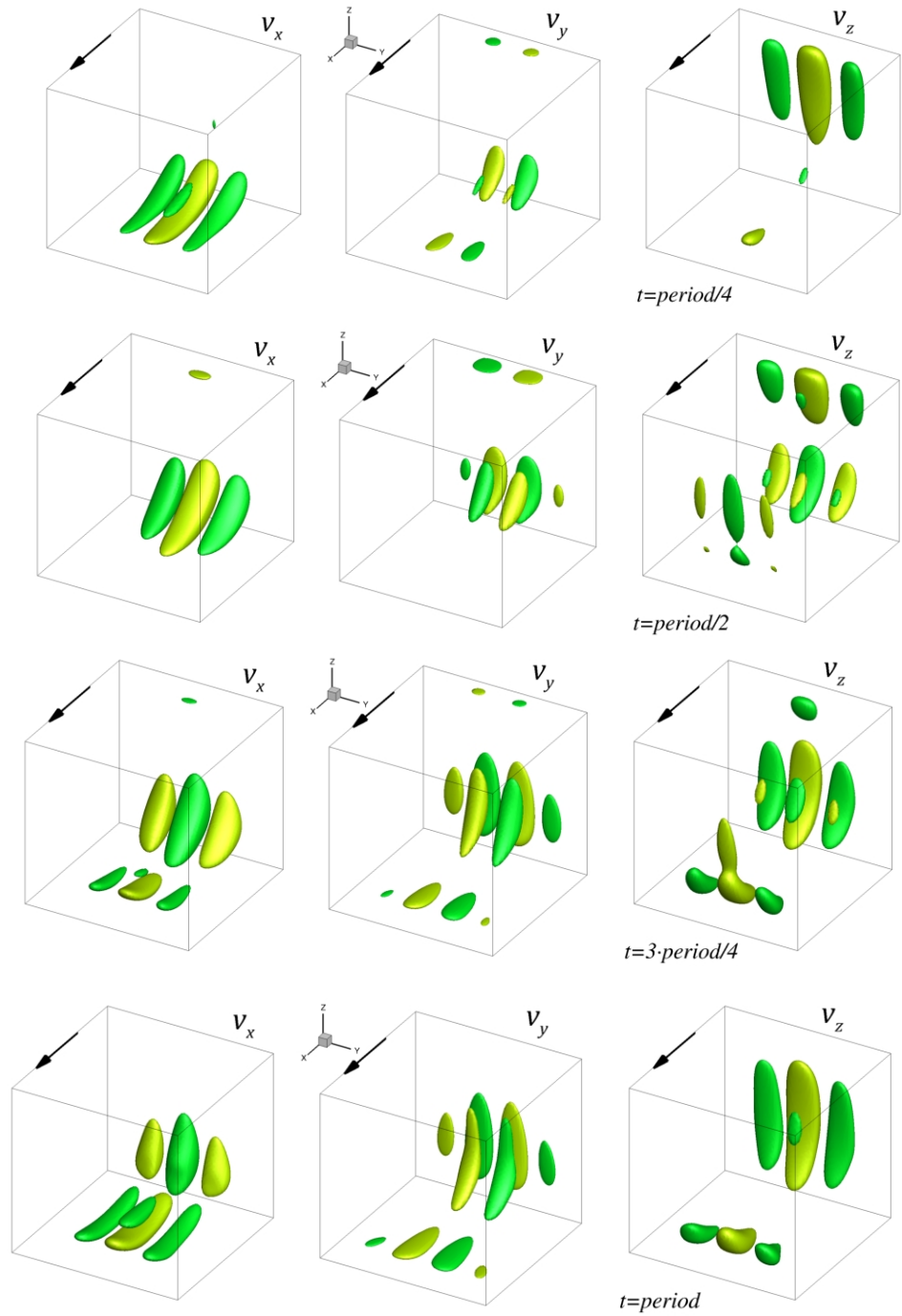

Figure 7. Isosurfaces of the leading velocity eigenvector at four equally distanced times along the oscillation period for the case $\alpha = 0$ (Animation 1). The values plotted are ±0.0015 for each velocity component. The minimal and maximal values are -0.00850 and 0.00568 for $v_x$, ±0.00569 for $v_y$, and -0.00411 and 0.00632 for $v_z$.

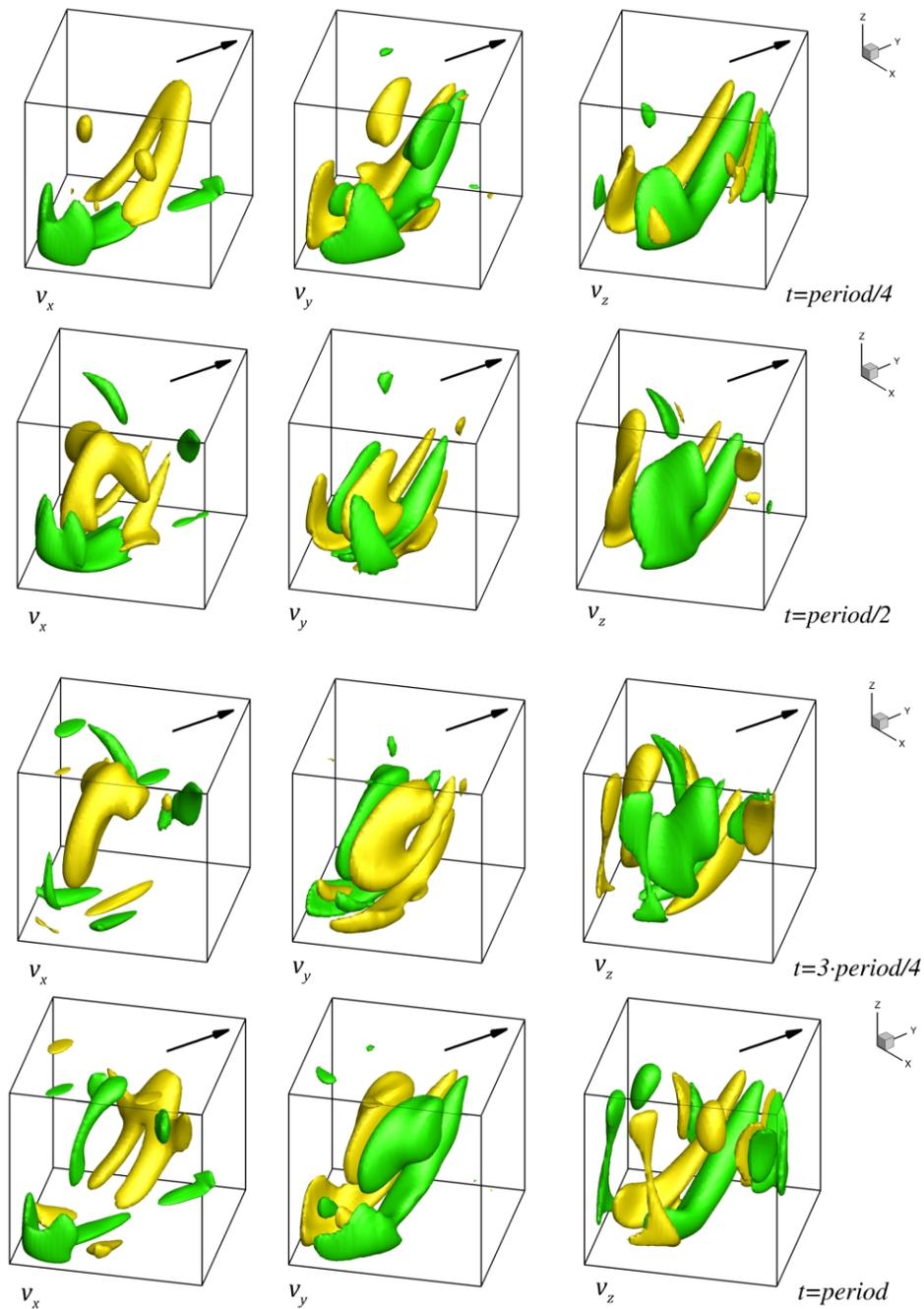

Figure 8. Isosurfaces of the leading velocity eigenvector at four equally distanced times along the oscillation period for the case $\alpha = 45°$ (Animation 2). The values plotted are ±0.0015 for each velocity component. The minimal and maximal values are -0.00342 and 0.00427 for $v_x$, -0.00775 and 0.00732 for $v_y$, and ±0.00823 for $v_z$.

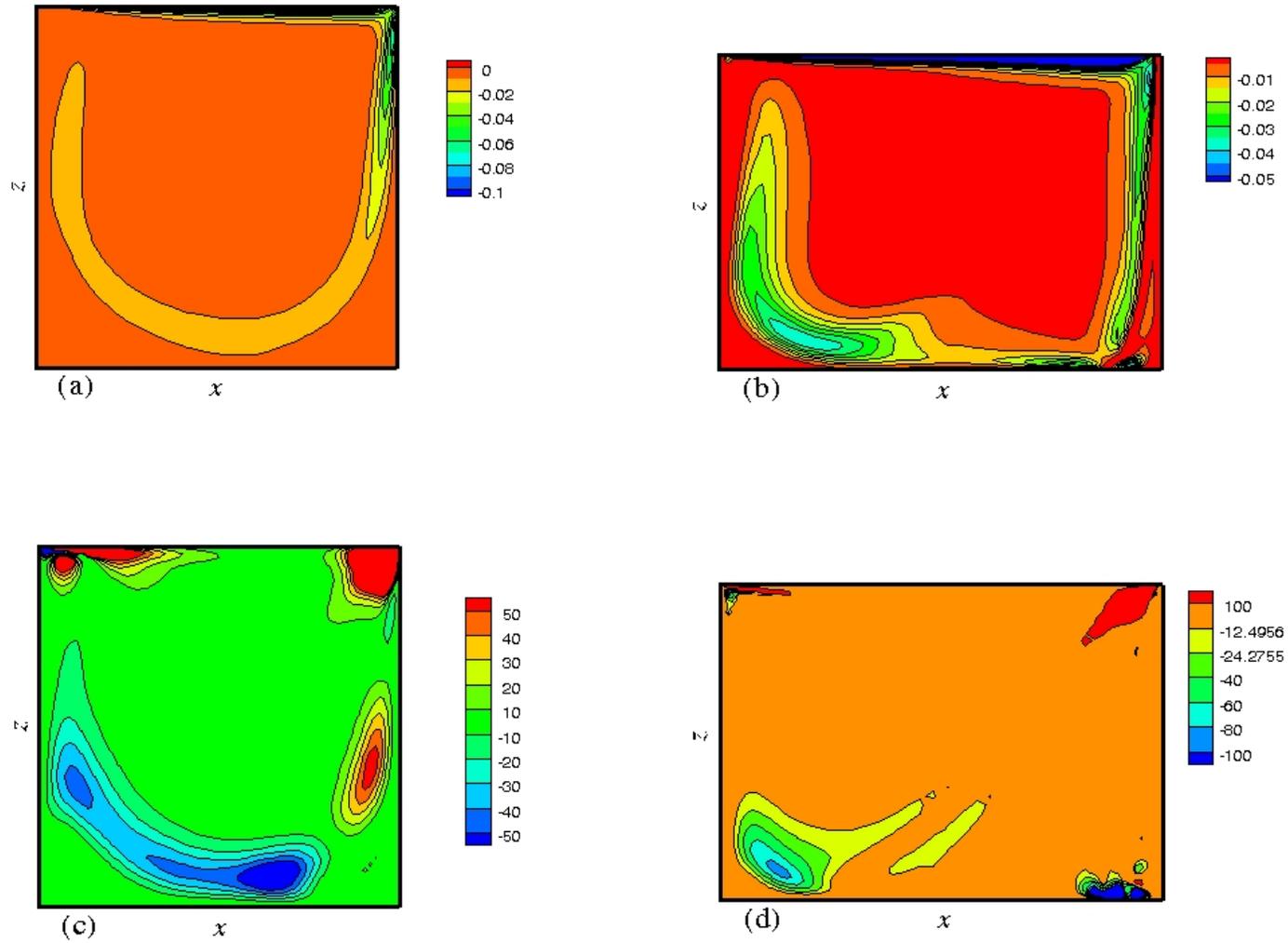

Figure 9. Function $-(\boldsymbol{r} \times \boldsymbol{v})^2$ of the Rayleigh stability criterion (a,b) and the quantity $B$ of the Bayly criterion (c,d) computed at the symmetry midplane for the slightly subcritical flows at (a) $a = 0$ and (b) $\alpha = 45°$.

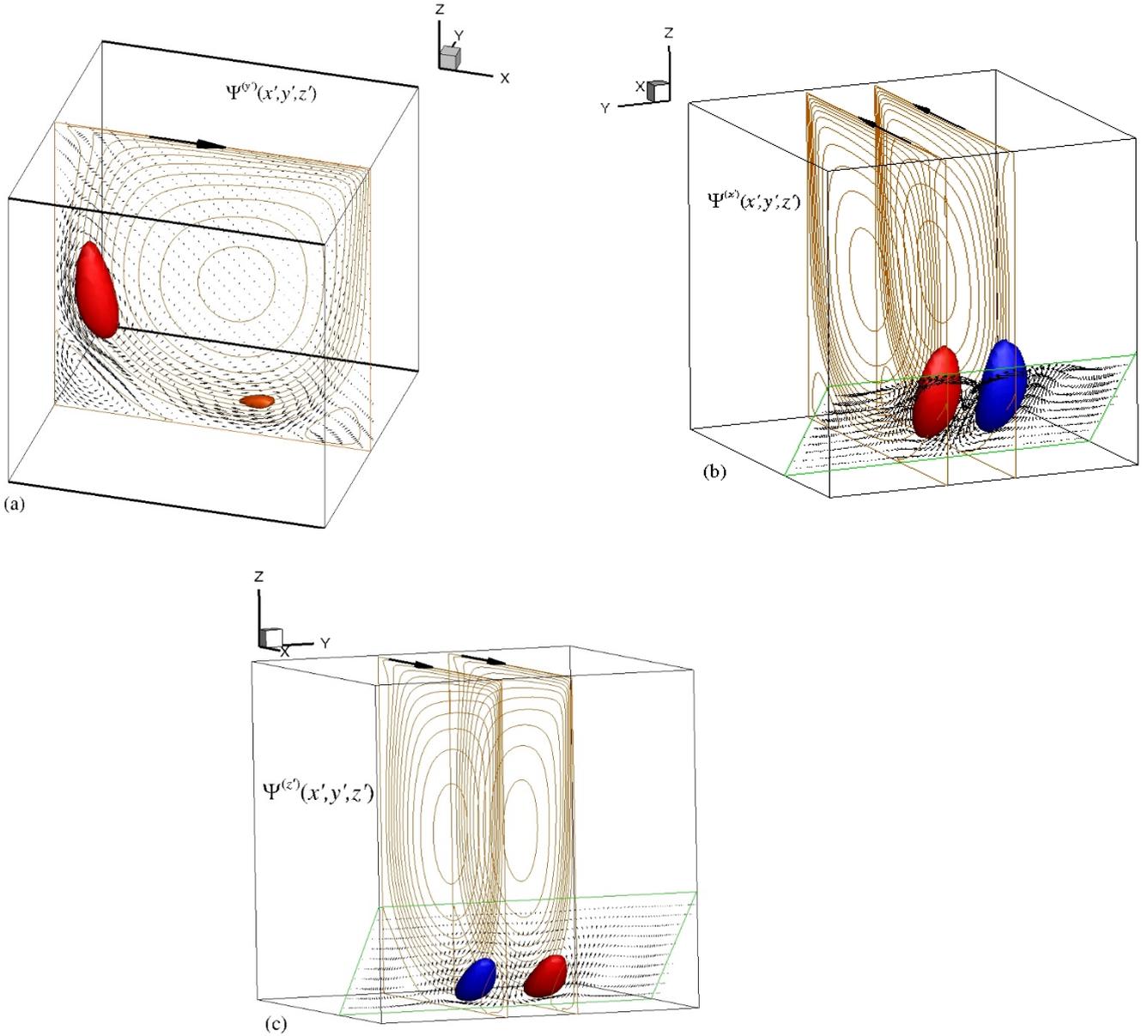

Figure 10. Visualization of the most unstable perturbation snapshots in axes rotated by Eq. (18), for the $\alpha = 0$ case. Brown streamlines correspond to the vector potential $\Psi^{(y')}$ of the base flow. The isosurfaces and the arrows show the vector potentials and the divergence-free velocity projections calculated for the perturbation snapshots (Animation 3). The isosurfaces are plotted for the levels -0.00023 and 0.00028 for $\Psi^{(y')}$, ±0.0003 for $\Psi^{(x')}$, and ±0.00022 for $\Psi^{(z')}$. The minimal and maximal values of the vector potentials are -0.0630 and 0.00417 for $\Psi^{(y')}$, -0.0142 and 0.01367 for $\Psi^{(x')}$, and ±0.00868 for $\Psi^{(z')}$.

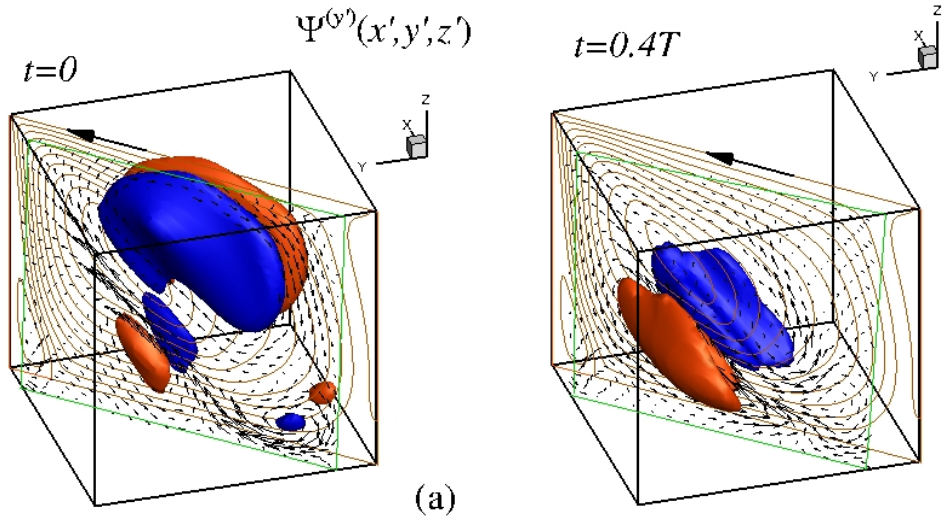

(a)

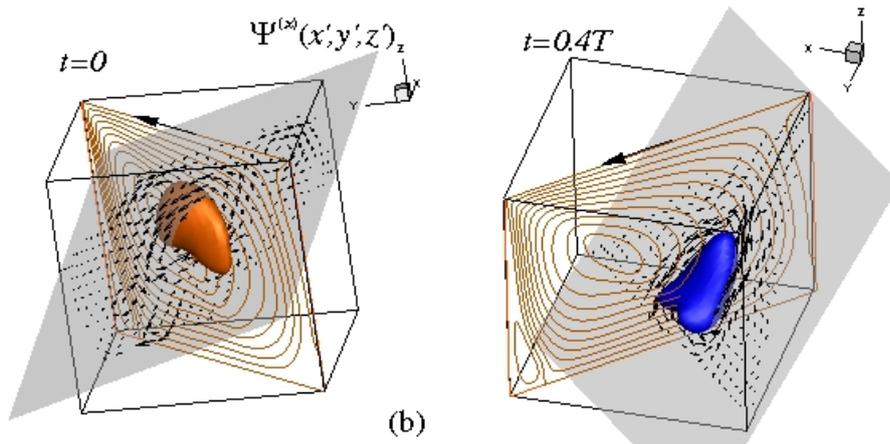

(b)

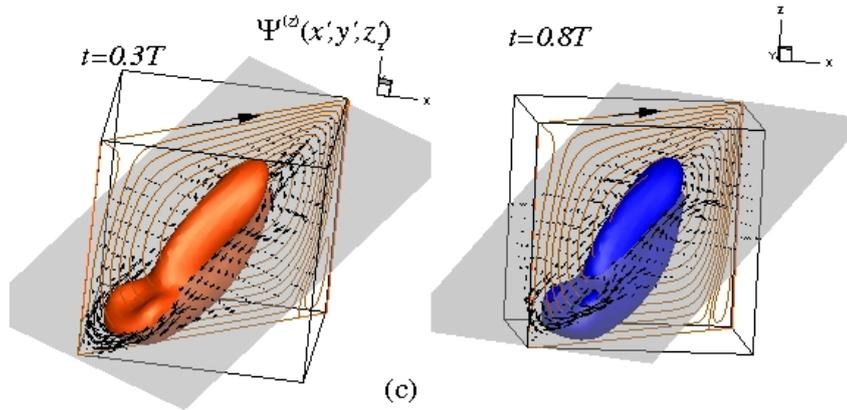

(c)

Figure 11. Visualization of the most unstable perturbation snapshots in axes rotated by Eq. (19), for the $\alpha = 45°$ case. Brown streamlines correspond to the vector potential $\Psi^{(y\prime)}$ of the base flow. The isosurfaces and the arrows show the vector potentials and the divergence-free velocity projections calculated for the perturbation snapshots (Animation 4). Gray color indicates planes in which the arrows are plotted. The isosurfaces are plotted for the levels $\pm 2.5 \times 10^{-5}$ for $\Psi^{(y\prime)}$, $\Psi^{(x\prime)}$, and $\Psi^{(z\prime)}$. The minimal and maximal values of the vector potentials are $-1.23 \times 10^{-4}$ and $5.42 \times 10^{-4}$ for $\Psi^{(y\prime)}$, $-2.15 \times 10^{-4}$ and $2.90 \times 10^{-4}$ for $\Psi^{(x\prime)}$, and $-3.94 \times 10^{-4}$ and $3.81 \times 10^{-4}$ for $\Psi^{(z\prime)}$.